\documentclass[aps,prb,superscriptaddress,amsfonts,amsmath,amssymb,showpacs,floatfix,reprint,longbibliography,footinbib]{revtex4-2}%

\usepackage{graphicx}
\usepackage{amsmath,amssymb}
\usepackage{dcolumn}
\usepackage{bm}
\usepackage{color}
\usepackage{hyperref}
\hypersetup{colorlinks=true,breaklinks,linkcolor=blue,urlcolor=blue,citecolor=blue}
\usepackage{xspace}
\usepackage{multirow}
\usepackage{physics}
\usepackage[version=4]{mhchem}

\newcommand{\cerhge}{\ce{CeRh6Ge4}}

\begin{document}

\title{From localized $4f$ electrons to anisotropic exchange interactions in ferromagnetic {\cerhge}}

\author{Shoichiro Itokazu}
\affiliation{Department of Physics, Okayama University, Okayama 700-8530, Japan}
\author{Akimitsu Kirikoshi}
\affiliation{Research Institute for Interdisciplinary Science, Okayama University, Okayama 700-8530, Japan}
\author{Harald O. Jeschke}
\email{jeschke@okayama-u.ac.jp}
\affiliation{Research Institute for Interdisciplinary Science, Okayama University, Okayama 700-8530, Japan}
\author{Junya Otsuki}
\email{j.otsuki@okayama-u.ac.jp}
\affiliation{Research Institute for Interdisciplinary Science, Okayama University, Okayama 700-8530, Japan}

\date{\today}

\begin{abstract}
{\cerhge} is a cerium-based ferromagnetic material exhibiting a quantum critical behavior under pressure.
We derive effective exchange interactions, using the framework of density functional theory combined with dynamical mean-field theory. 
Our results reveal that the nearest-neighbor ferromagnetic interaction along the $c$ axis is isotropic in spin space, leading to a formation of spin chains. On the other hand, the inter-chain coupling is highly anisotropic: The in-plane moment weakly interacts ferromagnetically in the $a$--$b$ plane to stabilize the ferromagnetic state, whereas the $z$-component couples antiferromagnetically, contributing to its destabilization.
The magnetic anisotropy of the interchain interactions as well as of the local $4f$ wavefunctions characterizes the magnetic properties underlying the ferromagnetic transition and the quantum critical behavior in {\cerhge}.
\end{abstract}

\maketitle

\section*{Introduction}
\label{sec:introduction}

Cerium based ferromagnets are materials of high fundamental interest. The highly localized $4f$ electrons of cerium can show Kondo lattice behavior where the localized spin is screened by conduction electrons and forms a Kondo singlet, leading to a correlated paramagnetic ground state~\cite{Hewson1993}. This behavior is controlled by the strength of the hybridization between the localized moment and the conduction electrons. In the case of weak hybridization, the localized $4f$ electrons can interact via the Ruderman–Kittel–Kasuya–Yosida (RKKY) interaction which is mediated by conduction electrons. The result can be magnetically ordered states.

A particularly interesting possibility in the case of cerium magnetism is a ferromagnetic ground state which is realized in a small but growing number of compounds such as \ce{CeRuPO}~\cite{Krellner2007}, \ce{CeRu2Al2B}~\cite{Baumbach2012} and \ce{CePt}~\cite{Larrea2005}. The fact that ferromagnetic ordering temperatures $T_{\rm C}$ are typically small means that there often are experimentally accessible tuning parameters like pressure that allow suppression of $T_{\rm C}$ to zero temperature, providing access to a quantum phase transition instead of the usual thermal phase transition. The properties of materials near such a quantum critical point are highly nontrivial and interesting~\cite{Brando2016}. Ferromagnetic quantum critical points are often first order
as in \ce{UGe2}~\cite{Huxley2001} or in \ce{UCoAl}~\cite{Aoki2011}. Here, we plan to study, \ce{CeRh6Ge4} which is a rare example of a Kondo lattice system with ferromagnetic order at ambient pressure and a second-order ferromagnetic quantum critical point that is accessible at elevated pressures.

{\cerhge} exhibits a ferromagnetic transition at $T_\textrm{C}=2.5$\,K~\cite{Matsuoka2015}.
This compound has attracted significant interest due to the emergence of a ferromagnetic quantum critical point (QCP) under pressure~\cite{Kotegawa2019,Shen2020}.
To realize a QCP associated with a ferromagnetic transition, magnetic anisotropy plays a key role~\cite{Shen2020}, as it helps to avoid a first-order transition.
Theoretical investigations into the realization of a ferromagnetic QCP have subsequently explored the absence of inversion symmetry~\cite{Kirkpatrick2020}, the influence of the spin-orbit coupling~\cite{Miserev2022}, and the role of nonsymmorphic symmetry~\cite{Shin2024}.
In this paper, we apply a recently developed first-principles method for calculating the magnetic susceptibility and derive low-energy effective interactions. We demonstrate that the exchange interations responsible for the ferromagnetic transition is highly anisotropic, stabilizing the in-plane ferromagneitc moment of $4f$ electrons.

\begin{figure}[tb]
    \centering
    \includegraphics[width=0.8\linewidth]{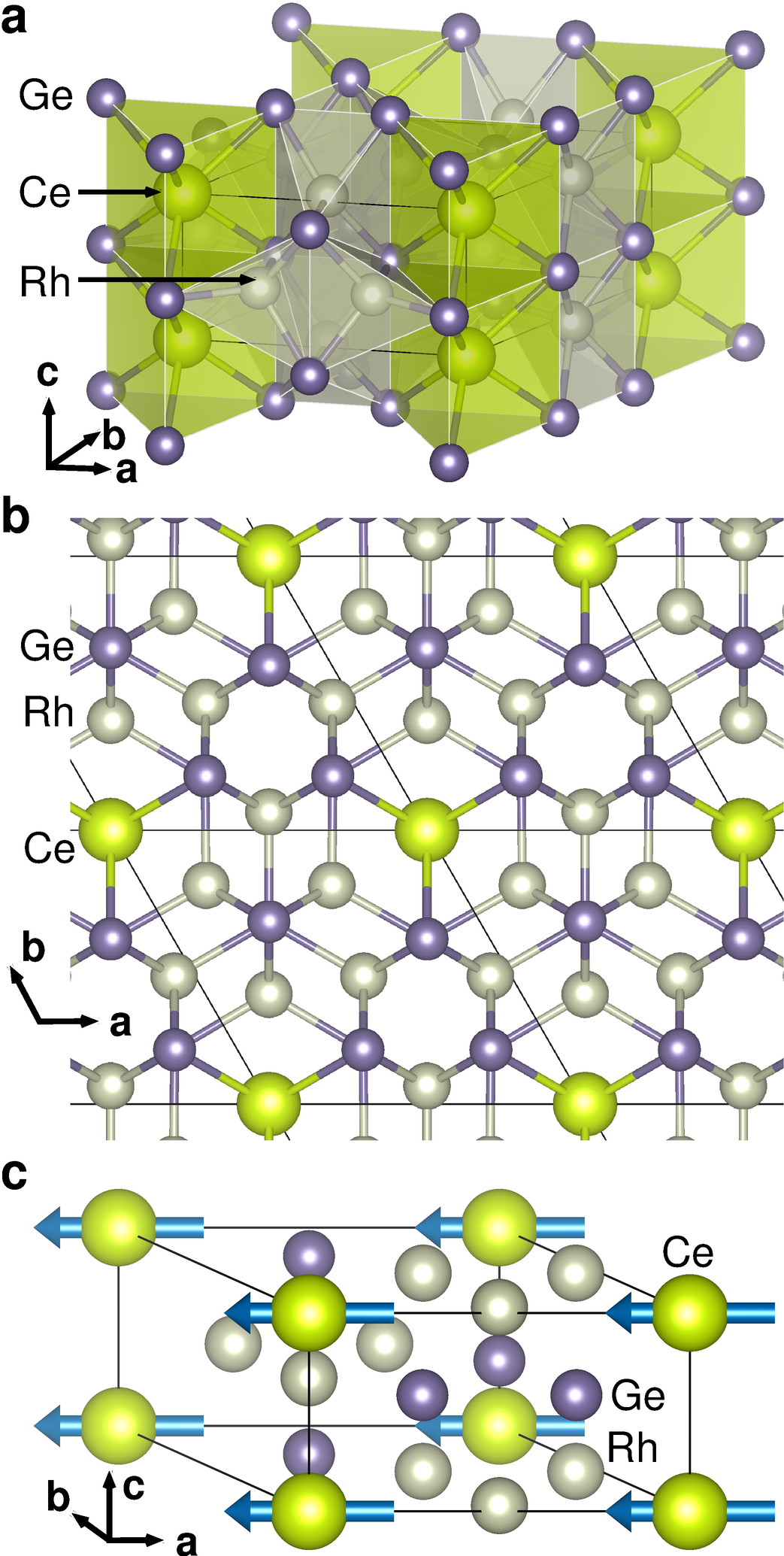}
    \caption{\textbf{The crystal structure of {\cerhge}.} \textbf{a} Side view showing the coordination of Ce and Rh. \textbf{b} View along $c$ showing the symmetry of the Ce site. \textbf{c} Spin configuration proposed in neutron scattering and $\mu$SR experiments~\cite{Shu2021}.}
    \label{fig:crystal}
\end{figure}

{\cerhge} crystallizes in a \ce{LiCo6P4}-type structure with space group $P\bar{6}m2$ (No.~187)~\cite{Vossinkel2012} as illustrated in Fig.~\ref{fig:crystal}.
The Ce atoms occupy the $1a$ site with a three-fold rotational symmetry, corresponding to the point group $D_{3h}$.
A clear anomaly in the specific heat indicates a phase transition at $T=T_\textrm{C}=2.5$\,K~\cite{Matsuoka2015}.
Magnetization measurements confirm the onset of ferromagnetism, with a spontaneous moment of $0.34\,\mu_\textrm{B}/\textrm{Ce}$.
Neutron scattering and $\mu$SR experiments reveal that the ordered moment lies within the $a$-$b$ plane as shown in Fig.~\ref{fig:crystal}c~\cite{Shu2021}.
In the paramagnetic state, the magnetic susceptibility follows the Curie-Weiss law with an effective moment of $2.35\,\mu_\textrm{B}/\textrm{Ce}$.
The electronic and magnetic properties of {\cerhge} have been further explored using angle-resolved photoemission spectroscopy (ARPES)~\cite{Wu2021},
quantum oscillation measurements~\cite{Wang2021},
chemical substitution at the Ce site~\cite{Xu2021} and the Ge site~\cite{Zhang2022}, and
thermopower measurements~\cite{Thomas2024}.

Theoretically, two contrasting approaches exist for describing $4f$ electron magnetism, depending on whether the $4f$ electrons are treated as itinerant or localized.
In the case of {\cerhge}, quantum oscillation measurements have shown that the Fermi surface is well described by models assuming localized $4f$ electrons~\cite{Wang2021}.
Consistently, ARPES measurements have also detected signatures characteristic of a localized $4f$ state~\cite{Wu2021}.
These experimental findings indicate that the localized picture provides a suitable starting point for understanding the electronic and magnetic properties of {\cerhge}.

We employ a method based on dynamical mean field theory combined with density functional theory (DFT+DMFT).
Our calculation procedure which successfully reproduced the antiferro-quadrupolar ordering in \ce{CeB6}~\cite{Otsuki2024} is the following.
First, we perform a DFT calculation in which the $4f$ electrons are treated as itinerant. Based on this electronic structure, we introduce the local Coulomb repulsion among the $4f$ electrons via DMFT, leading to $4f$ electrons with localized nature. We then calculate the momentum-dependent multipolar susceptibilities associated with the local $4f$ degrees of freedom. By identifying the divergence in the susceptibility, we determine the transition temperature and the corresponding order parameter.
Using this approach, we will demonstrate that the ferromagnetic transition in {\cerhge} can be reproduced within the localized $4f$ electron framework.


\section*{Results}
\begin{figure*}[t]
    \includegraphics[width=\linewidth]{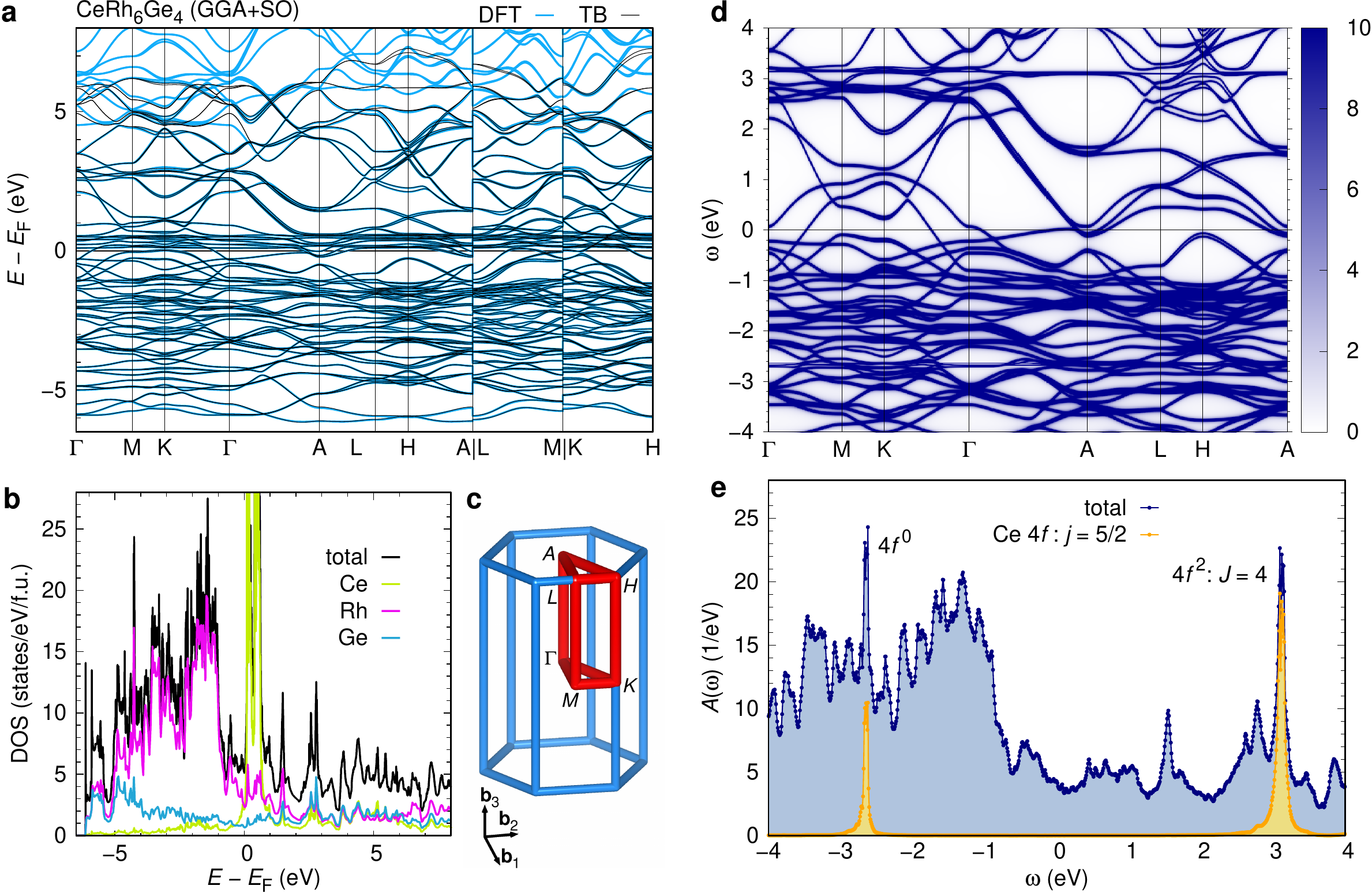}
    \caption{\textbf{Electronic structure of {\cerhge}.} \textbf{a} Fully relativistic DFT bandstructure (blue) with a 108 band tight binding fit (black). \textbf{b} Corresponding density of states of {\cerhge}. The maximum of the very sharp Ce $4f$ density of states of 76.2 states/eV/f.u. is not shown. \textbf{c} Brillouin zone of {\cerhge} with the high symmetry paths shown in \textbf{a}. 
    \textbf{d} Single-particle excitation spectrum $A(\bm{k},\omega)$ in DFT+DMFT calculated at $T=0.01$\,eV. \textbf{e} Corresponding $\bm{k}$-summed spectrum $A(\omega)$.}
    \label{fig:band_merged}
\end{figure*}

\noindent
{\bf Electronic structure}\\
Figure~\ref{fig:band_merged}a shows the electronic band structure calculated with fully relativistic density functional theory calculations using the full potential local orbital (FPLO) basis~\cite{Koepernik1999} in combination with a generalized gradient approximation exchange correlation functionals~\cite{Perdew1996}. 
We used the crystal structure reported in Ref.~\cite{Vossinkel2012}.
The lattice parameters are $a=7.154$\,{\AA} and $c=3.855$\,{\AA}. The path along the high symmetry points of the hexagonal space group of {\cerhge} is shown in Fig.~\ref{fig:band_merged}c.
Fig.~\ref{fig:band_merged}b shows the species resolved densities of states.
We use projective Wannier functions within FPLO~\cite{Eschrig2009,Koepernik2023} to obtain a tight binding model with 108 orbitals (including spin degrees of freedom) consisting of Ce $4f$, Ce $5d$, Rh $4d$, and Ge $4p$. Comparison of black and blue lines in Fig.~\ref{fig:band_merged}a indicate that the fit is excellent in a wide energy range around the Fermi level.
The maximum hybridization strengths between Ce $4f$ and Rh $4d$ and between Ce $4f$ and Ge $4p$ are 0.10\,eV and 0.15\,eV respectively, while the direct $4f$--$4f$ hopping is at most 2.9\,meV.

We performed magnetic calculations within LDA and GGA.
The quantization axis of the magnetic moment was chosen parallel to the $a$ axis or the $c$ axis.
In both cases, the total energy $E_\textrm{tot}$ takes minimum at $m=0$. This result demonstrates that the ferromagnetic state is not reproduced within LDA and GGA.

\begin{figure}[t]
    \centering
    \includegraphics[width=\linewidth]{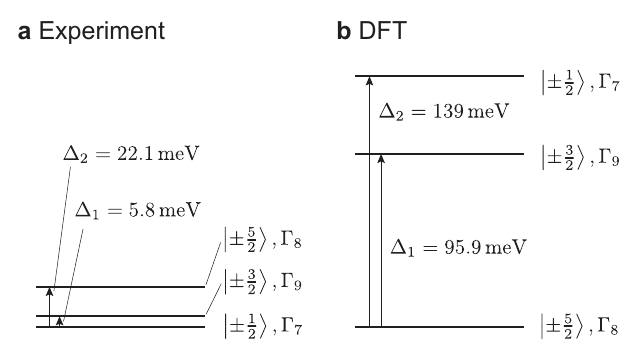}
    \caption{\textbf{CEF level schemes of $4f$ electrons in {\cerhge}.} \textbf{a} Scheme derived from experiment~\cite{Shu2021}. \textbf{b} CEF level scheme obtained by our DFT calculations.}
    \label{fig:CEF}
\end{figure}

Figure~\ref{fig:CEF} compares the crystalline electric field (CEF) energy levels of $4f$ electrons obtained from GGA calculations and experiment.
The local symmetry at the Ce site is described by the point group $D_{3h}$ under which the $j=5/2$ manifold splits into three Kramers doublets: $\ket{\pm1/2}$, $\ket{\pm3/2}$, and $\ket{\pm5/2}$.
Experimentally, the ground state is $\ket{\pm1/2}$, with the first and second excited states being $\ket{\pm3/2}$ and $\ket{\pm5/2}$, respectively~\cite{Shu2021}. The excitation energies are $\Delta_1=5.8$\,meV and $\Delta_2=22.1$\,meV, as determined from the temperature dependence of the magnetic susceptibility.
In contrast, our DFT calculations yield the opposite level ordering. Furthermore, the energy scale of the splitting is around 100\,meV, which is 5 times larger than the experimental one. 
The CEF levels in DFT do not reproduce the experimental magnetic anisotropy.
To address this discrepancy, we incorporate the experimentally determined CEF level scheme directly into our tight-binding Hamiltonian.

We treat the Coulomb repulsion in the $4f$ orbitals within the DFT+DMFT framework (see Methods section for details)~\cite{Georges1996,Held2001,Kotliar2006}.
We exclude the total angular momentum $j=7/2$ states of the $4f$ orbitals, which lie approximately $0.3$\,eV above the $j=5/2$ states due to the spin-orbit coupling.
The DFT+DMFT calculations are therefore applied to the remaining 100 orbitals.
Figure~\ref{fig:band_merged}d shows the single-particle excitation spectrum $A(\bm{k},\omega)$ calculated with parameters
$U=6.92$\,eV, $J_\textrm{H}=0.8$\,eV, and $\epsilon_f=-2.7$\,eV.
These values were chosen to ensure consistency between our results for $A(\bm{k},\omega)$ and experiments.
The Hund's coupling $J_\textrm{H}$ is adopted from Ref.~\cite{Locht2016}.
The $4f^0$ peak in $A(\bm{k},\omega)$ has been observed at $\omega=-\Delta_{-}$ with $\Delta_{-}=2.7$\,eV in photoemmision experiments~\cite{Wu2021}.
The lowest excitation energy from the $4f^1$ to the $4f^2$ configurations, $\Delta_{+}=3.1$\,eV, has been reported for elemental Ce~\cite{Herbst1978} and confirmed by BIS experiments~\cite{Lang1981}.
We determined the values of $U$ and $\epsilon_f$ to reproduce both $\Delta_{-}$ and $\Delta_{+}$ in our spectrum.
The resultant spectrum in Fig.~\ref{fig:band_merged}e exhibits no $4f$ spectral weight at the Fermi level, meaning that the $4f$ electrons are fully localized. The conduction band near the Fermi level primarily consists of Rh-$4d$ and Ge-$4p$ states.
\\

\noindent
{\bf Magnetic properties}\\
To investigate the phase transitions in {\cerhge}, we calculate the momentum-dependent static susceptibility, defined as
\begin{align}
    \chi_{m_1 m_2 m_3 m_4}(\bm{q})
    &= \frac{1}{N} \sum_{ij}
    e^{-i\bm{q} \cdot (\bm{R}_i - \bm{R}_j)}
    \nonumber \\
    &\times
    \int_0^{\beta} d\tau \langle O_{i, m_1 m_2}(\tau) O_{j, m_3 m_4} \rangle,
    \label{eq:chiq_1234}
\end{align}
where $m=+5/2, \cdots, -5/2$ denotes the $z$ component of $j=5/2$ orbitals, $N$ is the number of lattice sites, the argument $\tau$ indicates the imaginary-time evolution in the Heisenberg picture, and the operator $O_{i,mm'}$ is given by
\begin{align}
    O_{i, m m'} = f_{im}^{\dag} f_{im'}.
\end{align}
We evaluate $\chi_{m_1 m_2 m_3 m_4}(\bm{q})$ defined in Eq.~\eqref{eq:chiq_1234} using the strong-coupling-limit (SCL) formula (see Methods section for details).
The momentum-dependent susceptibilities in the SCL formula are given by
\begin{align}
    \hat{\chi}(\bm{q}) = \left[ \hat{\chi}_\textrm{loc}^{-1} - \hat{I}(\bm{q}) \right]^{-1},
    \label{eq:chi_SCL}
\end{align}
where all quantities with hat are matrices indexed by the combined indices $(m_1 m_2)$ and $(m_3 m_4)$.
$\hat{\chi}_\textrm{loc}$ denotes the local susceptibility calculated from the atomic system, and $\hat{I}(\bm{q})$ represents the intersite exchange interactions.

\begin{figure*}[t]
    \centering
    \includegraphics[width=\linewidth]{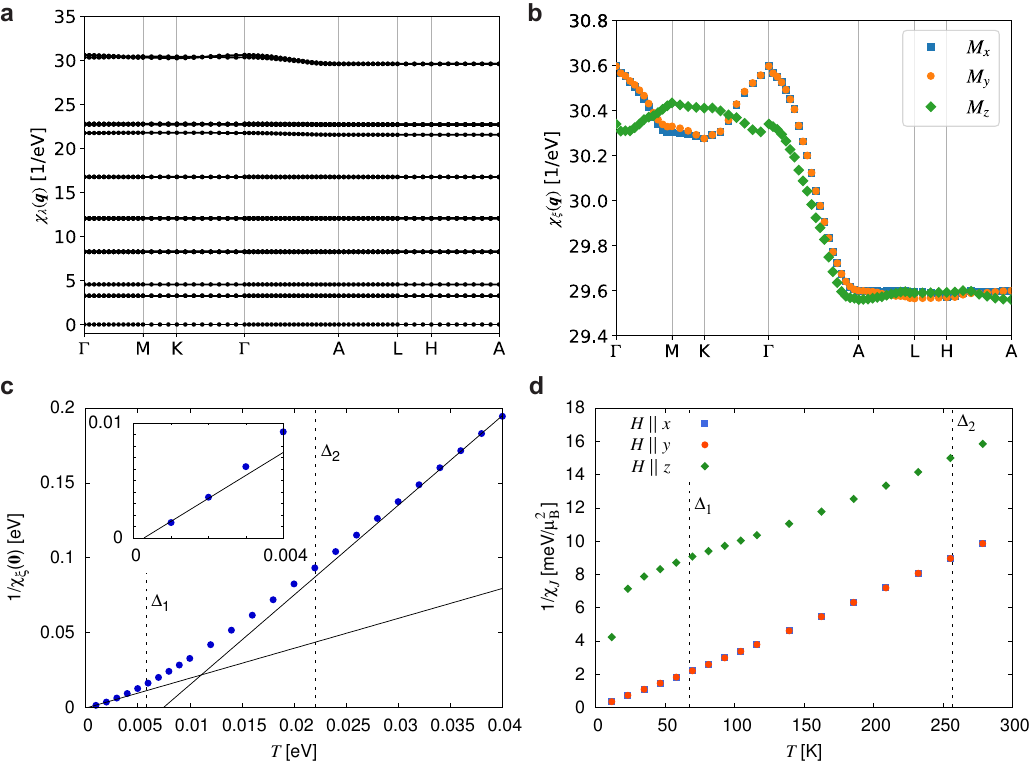}
    \caption{\textbf{The susceptibilities in momentum space and their temperature dependence. a.} The eigenvalues of the susceptibility matrix, $\chi_{\lambda}(\bm{q})$ computed at $T=0.01$\,eV. \textbf{b} The magnetic susceptibility $\chi_{\xi}(\bm{q})$ within $\ket{\pm1/2}$ states. \textbf{c} The temperature dependence of the inverse of the ferromagnetic susceptibility, $1/\chi_{\xi}(\bm{0})$ with $\xi=x, y$. The vertical dashed lines indicate the energy of the CEF level splitting $\Delta_1$ and $\Delta_2$ given in Fig.~\ref{fig:CEF}a. The solid lines show the fitting by the high-$T$ and low-$T$ expressions in Eqs.~\eqref{eq:chi_high} and \eqref{eq:chi_low}, respectively. The inset shows a zoom-up of the low-temperature region. \textbf{d} The temperature dependence of the inverse of the susceptibility $\chi_J$ that corresponds to experiments.}
    \label{fig:chi_merged}
\end{figure*}

\begin{table}[tb]
    \centering
    \caption{\textbf{Classification of eigenmodes $\chi_{\lambda}(\bm{q})$ of the susceptibility.} The order follows the result in Fig.~\ref{fig:chi_merged}a. ``States'' shows the CEF states related to the fluctuations. ``Irrep'' shows the irreducible representations of the fluctuations in point group $D_{3h}$, where the totally symmetric representation is expressed by $A_1^{\prime}$, and the superscript $+$ and $-$ indicate the time-reversal even and odd, respectively.}
    \label{tab:eigenmodes}
    \begin{tabular}{cll}
        \hline
        Dimension & States & Irrep \\
        \hline
        3 & $\ket{\pm1/2}$ & $A_2^{\prime -}\oplus E^{\prime\prime-}$ \\
        8 & $\ket{\pm1/2}$, $\ket{\pm3/2}$ & $E^{\prime\pm}\oplus E^{\prime\prime\pm}$ \\
        1 & all & $A_{1}^{\prime+}$ \\
        3 & $\ket{\pm3/2}$ & $A_1^{\prime\prime-}\oplus A_2^{\prime-}\oplus A_2^{\prime\prime-}$ \\
        8 & $\ket{\pm1/2}$, $\ket{\pm5/2}$ & $A_1^{\prime\prime\pm}\oplus A_2^{\prime\prime\pm}\oplus E^{\prime\pm}$ \\
        8 & $\ket{\pm3/2}$, $\ket{\pm5/2}$ & $E^{\prime\pm}\oplus E^{\prime\prime\pm}$ \\
        1 & all & $A_{1}^{\prime+}$ \\
        3 & $\ket{\pm5/2}$ & $A_2^{\prime-}\oplus E^{\prime\prime-}$ \\
        1 & all & $A_{1}^{\prime+}$ \\
        \hline
    \end{tabular}
\end{table}

Figure~\ref{fig:chi_merged}a shows the eigenvalues $\chi_{\lambda}(\bm{q})$ of the susceptibility matrix $\hat{\chi}(\bm{q})$.
A total of 36 eigenvalues are classified into 9 groups.
The number of modes in each group and the relevant CEF levels are summarized in Table~\ref{tab:eigenmodes}.
We determined the symmetry of the eigenmodes by analyzing the eigenvectors.
The group with the largest susceptibility corresponds to fluctuations within the CEF ground-state doublet $\ket{\pm1/2}$. This Kramers doublet carries magnetic degrees of freedom, which are classified into $A_2^{\prime-}$ and $E^{\prime\prime-}$ representations in the point group $D_{3h}$.
Groups with dimension 8 involve hybridization between two CEF doublets.
The corresponding density operators, such as $f_{1/2}^{\dag} f_{3/2}$, can be classified into pairs of the electric ($+$) and magnetic ($-$) multipole operators~\cite{Kusunose2023}.
Groups with dimension 1 consist of the totally symmetric representation $A_1^{\prime+}$. Among them, the lowest fluctuation mode, which exhibits nearly zero susceptibility, corresponds to charge fluctuations. The remaining modes are higher-order electric multipoles, including quadrupole and hexadecapole.

We focus on the leading fluctuations, which originate from the CEF ground-state doublet $\ket{\pm1/2}$.
The corresponding eigenvectors of the susceptibility matrix are well described by the Pauli matrices acting within the $\ket{\pm1/2}$ subspace.
To capture the magnetic fluctuations, we introduce the magnetic dipole operator projected onto the $\ket{\pm1/2}$ states as
\begin{align}
    M_{i\xi} = \sum_{mm'} f_{im}^{\dag} (\hat{M}_{\xi})_{mm'} f_{im'},
\end{align}
where $\xi=x, y, z$ and
the matrix $\hat{M}_{\xi}$ is defined by
\begin{align}
    \hat{M}_x &= \frac{1}{\sqrt{2}}
    \begin{pmatrix}
        0 & 0 & 0 & 0 & 0 & 0 \\
        0 & 0 & 0 & 0 & 0 & 0 \\
        0 & 0 & 0 & 1 & 0 & 0 \\
        0 & 0 & 1 & 0 & 0 & 0 \\
        0 & 0 & 0 & 0 & 0 & 0 \\
        0 & 0 & 0 & 0 & 0 & 0
    \end{pmatrix},
    \\
    \hat{M}_y &= \frac{1}{\sqrt{2}}
    \begin{pmatrix}
        0 & 0 & 0 & 0 & 0 & 0 \\
        0 & 0 & 0 & 0 & 0 & 0 \\
        0 & 0 & 0 & -i & 0 & 0 \\
        0 & 0 & i & 0 & 0 & 0 \\
        0 & 0 & 0 & 0 & 0 & 0 \\
        0 & 0 & 0 & 0 & 0 & 0
    \end{pmatrix},
    \\
    \hat{M}_z &= \frac{1}{\sqrt{2}}
    \begin{pmatrix}
        0 & 0 & 0 & 0 & 0 & 0 \\
        0 & 0 & 0 & 0 & 0 & 0 \\
        0 & 0 & 1 & 0 & 0 & 0 \\
        0 & 0 & 0 & -1 & 0 & 0 \\
        0 & 0 & 0 & 0 & 0 & 0 \\
        0 & 0 & 0 & 0 & 0 & 0
    \end{pmatrix}.
\end{align}
The susceptibility corresponding to the fluctuation of $M_{i\xi}$ can be evaluated by 
\begin{align}
    \chi_{\xi}(\bm{q}) = \sum_{m_1 m_2 m_3 m_4} (\hat{M}_{\xi})_{m_1 m_2}^{\ast} \chi_{m_1 m_2 m_3 m_4}(\bm{q}) (\hat{M}_{\xi})_{m_3 m_4}.
    \label{eq:chi_xi}
\end{align}
Figure~\ref{fig:chi_merged}b shows $\chi_{\xi}(\bm{q})$ on the $\bm{q}$-path.
These quantities closely follow the leading eigenvalues $\chi_{\lambda}(\bm{q})$ presented in Fig.~\ref{fig:chi_merged}a, confirming that $\chi_{\xi}(\bm{q})$ effectively captures the dominant magnetic fluctuations.
The strongest fluctuations appear in $\chi_x(\bm{q})$ and $\chi_y(\bm{q})$ at $\bm{q}=\bm{0}$, corresponding to ferromagnetic fluctuations of the in-plane magnetic moment $M_x$ and $M_y$.
The difference between $\chi_x(\bm{0})$ and $\chi_y(\bm{0})$ is small (less than 1\%), and we focus on the $x$ component hereafter.
Furthermore, $\chi_x(\bm{q})$ is significantly enhanced in the $q_z=0$ plane ($\Gamma$--M--K--$\Gamma$) compared to the $q_z=1/2$ plane (A--L--H--A). This indicates that the $c$ axis bond favors ferromagnetic configuration.

Figure~\ref{fig:chi_merged}c shows the temperature dependence of the inverse of the ferromagnetic susceptibility at ${\bm q}=0$, $1/\chi_{x}(\bm{0})$.
In the high-temperature region ($T \gg \Delta_2$), the six states of $j=5/2$ are effectively degenerate due to thermal fluctuations.
As a result, $\chi_{\xi}(\bm{q})$ follows the Curie-Weiss law represented by
\begin{align}
    \chi_{x}^\textrm{high} = \frac{C_6}{T-\Theta},
    \label{eq:chi_high}
\end{align}
where the Curie constant $C_6$ is evaluated under the assumption of no CEF splitting as 
\begin{align}
    C_6 \equiv \frac{1}{6} \sum_{m} (\hat{M}_{\xi}^2)_{mm} = \frac{1}{6}.
\end{align}
A fit to the numerical data yields the Curie-Weiss temperature of $\Theta=7.5$\,meV.
In the low-temperature region ($T \ll \Delta_1$), on the other hand,
only the lowest CEF doublet is thermally occupied.
In this limit, $\chi_{\xi}(\bm{q})$ can be represented by
\begin{align}
    \chi_{x}^\textrm{low} = \frac{C_2}{T-T_\textrm{C}},
    \label{eq:chi_low}
\end{align}
and the Curie constant $C_2$ is evaluated only with the $\ket{\pm1/2}$ states as
\begin{align}
    C_2 \equiv \frac{1}{2} \sum_{m=\pm 1/2} (\hat{M}_{\xi}^2)_{mm} = \frac{1}{2}.
\end{align}
Fitting the low-temperature data gives a ferromagnetic transition temperature of $T_\textrm{C} = 0.28\,\textrm{meV} \approx 3.2$\,K. It is important to note that this first principle result for $T_\textrm{C}$ is very close to the experimental value of $T_\textrm{C} =2.5$\,K. 
This overestimation by approximately 0.7\,K or 30{\%} is consistent with previous findings for \ce{CeB6}, where the SCL method similarly overestimated the transition temperature of the antiferro-quadrupolar ordering~\cite{Otsuki2024}.

We note that the magnetic susceptibility $\chi_{\xi}$ defined in Eq.~\eqref{eq:chi_xi} differs from the quantity measured in experiments.
The experimentally measured magnetic susceptibility $\chi_J$ is defined by 
\begin{align}
    \chi_J = \left. \pdv{\expval{m}}{H} \right|_{H\to0},
\end{align}
under the Zeeman field $\mathcal{H}_Z=-mH$.
For a magnetic field parallel to the $z$ axis, the magnetic moment operator $m$ is given by
$m=-g_J \mu_\mathrm{B} j_z$,
where $j_z$ is the $z$-component of the total angular momentum operator with $j=5/2$ and $g_J$ is the Land\'e $g$-factor, which is given by $g_J=6/7$ for the Ce$^{3+}$ ion.
$\chi_J$ can be computed from $\chi_{m_1 m_2, m_3 m_4}(\bm{q})$ as in Eq.~(2).
The explicit form for $\chi_J$ is thus given by
\begin{align}
    \chi_J &= (g_J \mu_\mathrm{B})^2 
    \notag \\
    &\times \sum_{m_1 m_2 m_3 m_4} (j_z)_{m_1 m_2}^{\ast} \chi_{m_1 m_2 m_3 m_4}(\bm{0}) (j_z)_{m_3 m_4}.
    \label{eq:chi_J}
\end{align}
The cases with the magnetic field parallel to the $x$- and $y$-axes are computed in a similar manner.
Figure~\ref{fig:chi_merged}d shows the temperature dependence of $1/\chi_J$.
The overall behavior, including the anisotropy and the strong $T$-dependence for $H \parallel z$ below $T\simeq \Delta_1$, is consistent with the experiments~\cite{Shu2021}.

\begin{figure}[!t]
    \centering
    \includegraphics[width=\linewidth]{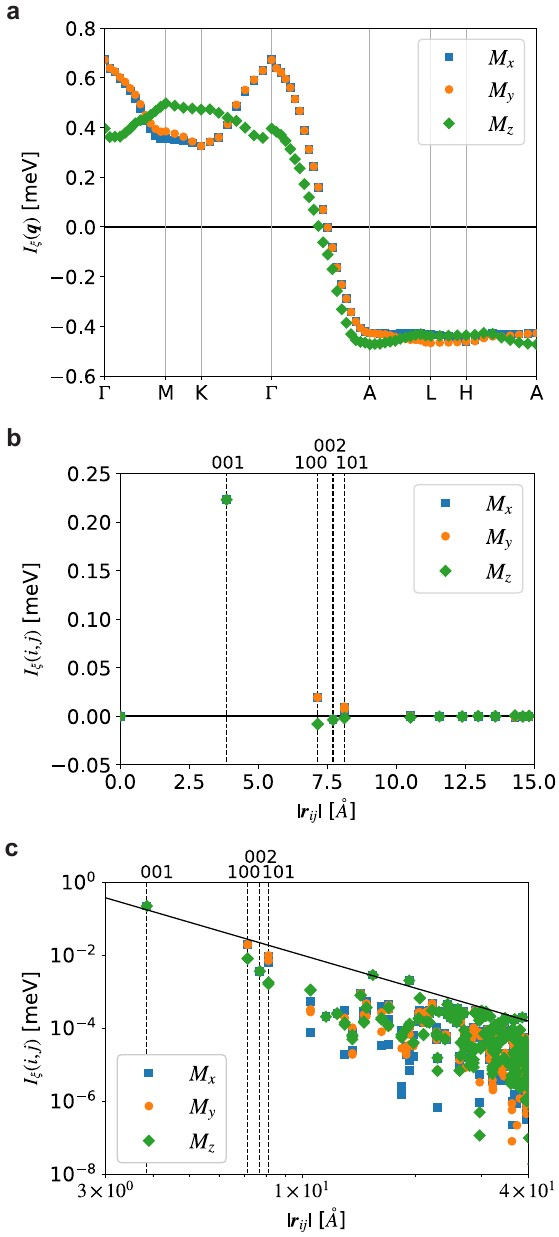}
    \caption{\textbf{Effective interactions between the magnetic moments in the $\ket{\pm1/2}$ states.}
    \textbf{a} The momentum-dependent effective interaction $I_{\xi}(\bm{q})$.
    \textbf{b} Linear-scale plot of the effective inter-site interaction $I_{\xi}(i, j)$ as a function of the distance $|\bm{r}_{ij}| = |\bm{R}_i - \bm{R}_j|$.  \textbf{c} A log-log plot with the line showing $I_{\xi}(i, j) \propto |\bm{r}_{ij}|^{-3}$. The vertical dashed lines show the Ce-Ce distances with the labels $n_1 n_2 n_3$ indicating $\bm{r}=n_1 \bm{a} + n_2 \bm{b} + n_3 \bm{c}$.}
    \label{fig:I}
\end{figure}

We now turn to the intersite interactions.
To isolate the effective interaction between $\ket{\pm1/2}$ states, we project the full momentum-dependent interaction $I_{m_1 m_2 m_3 m_4}(\bm{q})$ onto the dipole channel using the same transformation as in Eq.~\eqref{eq:chi_xi}. The resulting quantity $I_{\xi}(\bm{q})$ represents the exchange interaction within the ground-state doublet.
Figure~\ref{fig:I}a shows $I_{\xi}(\bm{q})$ along the representative $\bm{q}$-path.
Its $\bm{q}$-dependence is closely follows that of $\chi_{\xi}(\bm{q})$, in agreement with Eq.~\eqref{eq:chi_SCL}, which indicates that the $\bm{q}$-dependence of $\chi_{\xi}(\bm{q})$ originates entirely from $I_{\xi}(\bm{q})$.
The magnitude of the ferromagnetic interaction is approximately $0.7$\,meV, which is comparable to the ferromagnetic transition temperature $T_\textrm{C}$.

To gain real-space insight into the magnetic interactions, we computed $I_{\xi}(\bm{q})$ over the entire Brillouin zone and performed a Fourier transform to obtain the intersite interactions $I_{\xi}(i, j)$.
This quantity defines the effective Heisenberg Hamiltonian for the projected dipole moments $M_{i\xi}$, given by
\begin{align}
    \mathcal{H}
    = -\frac{1}{2} \sum_{ij\xi}
    M_{i\xi} I_{\xi}(i,j) M_{j\xi}.
    \label{eq:H_eff}
\end{align}
Figure~\ref{fig:I}b shows $I_{\xi}(i, j)$ as a function of intersite distance $|\bm{R}_i - \bm{R}_j|$.
Explicit values of $I_{\xi}(i, j)$ are presented in Table~\ref{tab:I}.
The strongest interaction occurs along the $c$ axis, corresponding to the nearest-neighbor sites. This bond exhibits a sizable ferromagnetic exchange that is isotropic in spin space.
The next-nearest-neighbor interactions, located in the $a$--$b$ plane, exhibit pronounced spin anisotropy. Specifically, the in-plane components $M_x$ and $M_y$ are coupled ferromagnetically, while the out-of-plane component $M_z$ shows antiferromagnetic coupling.
Figure~\ref{fig:I}c presents a log-log plot of $|I_{\xi}(i, j)|$, revealing a power-low decay $I_{\xi}(i, j) \sim |\bm{R}_i - \bm{R}_j|^{-3}$. This behavior is characteristic of the Ruderman--Kittel--Kasuya--Yosida (RKKY) interaction, confirming its role as the primary mechanism driving the magnetic ordering in {\cerhge}.

\begin{table}[tb]
    \centering
    \begin{tabular}{c|rr}
        \hline
        $n_1 n_2 n_3$ & $I_x$\,[meV] & $I_z$\,[meV] \\
        \hline
        001 & 0.223 & 0.223 \\
        100 & 0.019 & $-0.008$ \\
        002 & $-0.004$ & $-0.004$ \\
        101 & 0.009 & $-0.002$ \\
        \hline
    \end{tabular}
    \caption{\textbf{Values of intersite interactions $I_{\xi}(i, j)$ corresponding to Fig.~\ref{fig:I}b.} Only $\xi=x, z$ components are shown because of $I_x(i,j) \simeq I_y(i,j)$.}
    \label{tab:I}
\end{table}

\begin{figure*}[t]
    \centering
    \includegraphics[width=\linewidth]{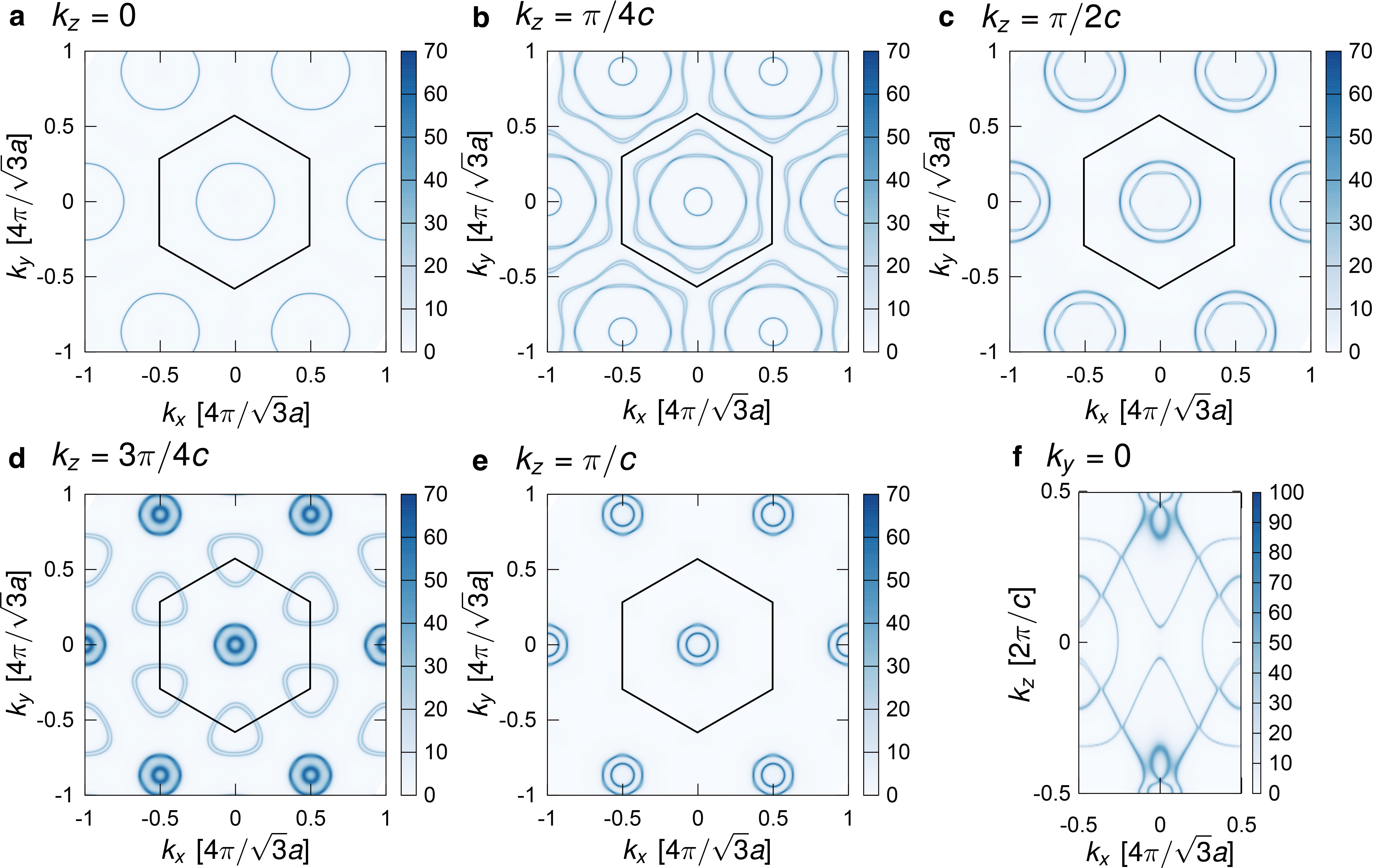}
    \caption{
    \textbf{Single-particle excitation spectrum $A(\bm{k},\omega)$ of {\cerhge} at $\omega=0$}. \textbf{a}--\textbf{e} The $k_z=n\pi/4c$ plane with $n=0, 1, 2, 3, 4$. The hexagon indicates the Brillouin zone. \textbf{f} The $k_{y}=0$ plane. A full Brillouin zone is shown.
    }
    \label{fig:FS}
\end{figure*}

\section*{Discussion and Conclusions}
\label{sec:discussion}

The magnetic structure obtained by our calculations is illustrated in Fig.~\ref{fig:crystal}c.
The effective interaction responsible for this structure can be described as follows.
By retaining interactions up to the second nearest neighbors in Eq.~\eqref{eq:H_eff}, we obtain the effective Heisenberg model of spin $S=1/2$ as
\begin{align}
    \mathcal{H} \simeq &-J_c \sum_{\langle ij \rangle \parallel c} \bm{S}_i \cdot \bm{S}_j
    \notag \\
    &- J_{ab} \sum_{\langle ij \rangle \perp c} \left[ S_i^x S_j^x + S_i^y S_j^y - r S_i^z S_j^z \right],
\end{align}
where the first and the second summations represent nearest-neighbor pairs along the $c$ axis and within the $a$--$b$ plane, respectively.
From the numerical results in Table.~\ref{tab:I}, we obtain $J_c=0.45$\,meV, $J_{ab}/J_c=0.087$, and $r=0.41$ (note that $J$'s are twice as large as $I_{\xi}$).
The first term represents the strong coupling of the Ce moments along the $c$ axis, leading to the formation of ferromagnetic chains.
These chains are aligned by weaker inter-chain interactions within the $a$--$b$ plane (the second term), which collectively stabilize the ferromagnetic configuration shown in Fig.~\ref{fig:crystal}c.
Although the in-plane components of the moment are coherently aligned, the $c$ axis component is influenced by geometrical frustration arising from the inter-chain antiferromagnetic coupling. This frustration suppresses out-of-plane magnetic order and is reflected in the momentum dependence of the interaction $I_z(\bm{q})$, whose maximum appears at the M point of the Brillouin zone [Fig.~\ref{fig:I}a].

In Ref.~\cite{Shen2020}, the strange-metal behavior around the QCP is analyzed by the Heisenberg Hamiltonian, assuming XXZ-type anisotropy with weak $S^z$ interaction. Our microscopic derivation of the effective Heisenberg model supports their starting point, which leads to a linear dispersion of the magnetic excitations.
Furthermore, the weak inter-chain coupling guarantees a second-order phase transition, enabling the emergence of the QCP under pressure~\cite{Shen2020}.

A direct calculation of QCP under external pressure requires substantial computational effort. 
As described by the Doniach phase diagram~\cite{Doniach1977}, the ordering temperature is suppressed as a result of competition between long-range order induced by the RKKY interactions and the paramagnetic heavy-Fermion state driven by the Kondo effect.
Our calculations using the atomic self-energy within the Hubbard-I approximation neglect the Kondo effect.
Consequently, applying the present calculations under pressure simply leads to an increase in the Curie temperature associated with the reduction of the lattice constant.
To estimate the Kondo temperature $T_\mathrm{K}$, we employed the continuous-time quantum Monte Carlo method~\cite{Gull2011} as an impurity solver. While we observed signs of a low-energy quasiparticle peak emerging below $T \simeq 1$\,meV, determining $T_\mathrm{K}$ and its pressure dependence with sufficient accuracy remains challenging.

Finally, we discuss the origin of the ferromagnetic transition in {\cerhge}.
The $\bm{q}$-dependence of the RKKY interaction is governed by the Fermi surface of conduction electrons.
Figure~\ref{fig:FS} displays the Fermi surface evaluated from $A(\bm{k},\omega)$ at $\omega=0$.
As shown in Figures~\ref{fig:FS}a--e, the Fermi surface is isotropic in $k_x$--$k_y$ plane.
The cut at the $k_y=0$ plane in Fig~\ref{fig:FS}f reveals a Fermi surface that spans the entire $k_z$ range, indicating a deformed cylindrical shape---characteristic of two-dimensional systems.
These features are consistent with a quantum oscillation measurement, which showed that the observed Fermi surface is consistent with the DFT results assuming localized $4f$ electrons~\cite{Wang2021}.
It is known that the two-dimensional free electrons exhibit a plateau in the static susceptibility $\chi_0(\bm{q})$ for $|\bm{q}|<2k_\textrm{F}$, followed by a decay for $|\bm{q}|>2k_\textrm{F}$, where $k_\textrm{F}$ is the Fermi wavenumber.
We suggest that the weakly three-dimensional cylindrical Fermi surface gives rise to the peak at the $\Gamma$ point in our numerical results.

\label{sec:summary}

In summary, we investigated the ferromagnetism in {\cerhge} using the DFT+DMFT method, treating the Ce $4f$ electrons as localized. 
The intersite exchange interactions, evaluated using the strong-coupling-limit (SCL) formula, revealed dominant ferromagnetic coupling along the $c$ axis, forming spin chains that are isotropic in spin space. In contrast, the inter-chain coupling is anisotropic: ferromagnetic for the in-plane moment and antiferromagnetic for the out-of-plane moment. These interactions give rise to a magnetic structure in which the magnetic moment is aligned within the $a$--$b$ plane.

The calculated transition temperature is in reasonable agreement with the experimental value, although the mean-field nature of the intersite interaction in DMFT tends to overestimate the transition temperature. 
Our results provide theoretical support 
for anisotropic exchange interactions between localized $4f$ moments,
offering a foundation for understanding the anomalous metallic behavior observed near the ferromagnetic QCP under pressure.

\section*{Methods}


\noindent
{\bf DFT+DMFT method}\\
The DFT+DMFT method incorporates local electronic correlations within a specific shell, building upon the DFT electronic structure~\cite{Georges1996,Held2001,Kotliar2006}.  
The single-particle Green's function matrix $\hat{G}(\bm{k}, \omega)$ is defined as
\begin{align}
    \hat{G}(\bm{k}, \omega)
    = \left[ (\omega+\mu)\hat{I} - \hat{H}(\bm{k}) - \hat{\Sigma}_\textrm{loc}(\omega) + \hat{\Sigma}_\textrm{DC} \right]^{-1}.
\end{align}
Here, all quantities with hats represent matrices in the Wannier-function basis.
$\hat{I}$ is the identity matrix, and $\mu$ is the chemical potential.
$\hat{H}(\bm{k})$ is the tight-binding Hamiltonian obtained from DFT, where the on-site energies of the $4f$ orbitals are adjusted to reproduce the experimental CEF energy level scheme.
$\hat{\Sigma}_\textrm{DC}$ is the double-counting correction that accounts for correlation effects already included in DFT. Following Ref.~\cite{Otsuki2024}, we represent it as
\begin{align}
    \hat{\Sigma}_\textrm{DC} = -\epsilon_f \ketbra{f},
\end{align}
where $\ketbra{f}$ is the projection operator onto the $4f$ subspace.
The parameter $\epsilon_f$ is determined along with the interaction parameters.

The local self-energy $\hat{\Sigma}(\omega)$ is defined on the $4f$ orbitals.
We employ the fully rotationally invariant Slater interactions, characterized by four Slater integrals $F_k$ with $k=0, 2, 4, 6$. These are converted from two intuitive parameters: the direct Coulomb interaction $U$ and the Hund's exchange coupling $J_\textrm{H}$~\cite{Anisimov1997}.
To compute $\hat{\Sigma}(\omega)$, we solve the atomic problem within the Hubbard-I approximation, which neglects hybridization with the conduction electrons.
These calculations were done using the open-source software \texttt{DCore} \cite{Shinaoka2021}, which is implemented with \texttt{TRIQS} \cite{Parcollet2015} and \texttt{DFTTools} \cite{Aichhorn2016} libraries.
The exact diagonalization of the atomic problem was solved using \texttt{pomerol} \cite{pomerol}.
The number of $\bm{k}$ points in the DFT+DMFT calculations is $24\times24\times42$.
\\

\noindent
\textbf{SCL formula for susceptibility}\\
We evaluate $\chi_{m_1 m_2 m_3 m_4}(\bm{q})$ defined in Eq.~\eqref{eq:chiq_1234} using the strong-coupling-limit (SCL) formula, which is derived from the Bethe-Salpeter (BS) equation within DMFT~\cite{Otsuki2019,Otsuki2024}.
Unlike the full BS equation, the SCL formula does not require computation of the vertex function, significantly simplifying the calculation.
When applied to \ce{CeB6}, the SCL formula yields quantitative agreement with the results obtained from the full BS equation~\cite{Otsuki2024}, demonstrating its reliability in systems with localized $4f$ electrons.

The momentum-dependent susceptibilities in the SCL formula are given by Eq.~\eqref{eq:chi_SCL}.
To evaluate $\hat{I}(\bm{q})$, we employ the two-pole approximation (SCL3)~\cite{Otsuki2019,Otsuki2024}, which captures the virtual excitations around the $4f^1$ configuration. We use excitation energies $\Delta_{+}=3.1$\,eV and $\Delta_{-}=2.7$\,eV, corresponding to excitations from $4f^1$ states to $4f^2$ and $4f^0$ configurations, respectively, in $A(\bm{k},\omega)$.

\section*{Data availability}


The data generated during the current study are available from the corresponding author (JO) on reasonable request.

\section*{Code Availability}

The DFT code used in this study is available from \url{https://www.fplo.de/}. 
The DMFT calculations for the single-particle excitation spectrum are performed using \texttt{DCore}, which is available from \url{https://github.com/issp-center-dev/DCore}. The calculations of the susceptibility and effective interactions are performed by \texttt{ChiQ}, which is now open in \url{https://github.com/issp-center-dev/ChiQ}.
\\

\section*{Author contributions}

The project was conceived by JO. SI performed the calculations with contributions from all authors. SI and JO analysed and discussed the results and drafted the manuscript with contributions from AK and HOJ. All authors revised the manuscript.

\section*{Competing interests}

The authors declare no competing interests.

\section*{Acknowledgments}

This work was supported by JSPS KAKENHI grants No. 21H01003, No. 21H01041, No. 23H04869, and No. 24H01668.
A part of the computation in this work has been done using the facilities of the Supercomputer Center, the Institute for Solid State Physics, the University of Tokyo (2024-Bb-0043).

\section*{References}

\bibliography{refs}

\begin{thebibliography}{40}%
\makeatletter
\providecommand \@ifxundefined [1]{%
 \@ifx{#1\undefined}
}%
\providecommand \@ifnum [1]{%
 \ifnum #1\expandafter \@firstoftwo
 \else \expandafter \@secondoftwo
 \fi
}%
\providecommand \@ifx [1]{%
 \ifx #1\expandafter \@firstoftwo
 \else \expandafter \@secondoftwo
 \fi
}%
\providecommand \natexlab [1]{#1}%
\providecommand \enquote  [1]{``#1''}%
\providecommand \bibnamefont  [1]{#1}%
\providecommand \bibfnamefont [1]{#1}%
\providecommand \citenamefont [1]{#1}%
\providecommand \href@noop [0]{\@secondoftwo}%
\providecommand \href [0]{\begingroup \@sanitize@url \@href}%
\providecommand \@href[1]{\@@startlink{#1}\@@href}%
\providecommand \@@href[1]{\endgroup#1\@@endlink}%
\providecommand \@sanitize@url [0]{\catcode `\\12\catcode `\$12\catcode `\&12\catcode `\#12\catcode `\^12\catcode `\_12\catcode `\%12\relax}%
\providecommand \@@startlink[1]{}%
\providecommand \@@endlink[0]{}%
\providecommand \url  [0]{\begingroup\@sanitize@url \@url }%
\providecommand \@url [1]{\endgroup\@href {#1}{\urlprefix }}%
\providecommand \urlprefix  [0]{URL }%
\providecommand \Eprint [0]{\href }%
\providecommand \doibase [0]{https://doi.org/}%
\providecommand \selectlanguage [0]{\@gobble}%
\providecommand \bibinfo  [0]{\@secondoftwo}%
\providecommand \bibfield  [0]{\@secondoftwo}%
\providecommand \translation [1]{[#1]}%
\providecommand \BibitemOpen [0]{}%
\providecommand \bibitemStop [0]{}%
\providecommand \bibitemNoStop [0]{.\EOS\space}%
\providecommand \EOS [0]{\spacefactor3000\relax}%
\providecommand \BibitemShut  [1]{\csname bibitem#1\endcsname}%
\let\auto@bib@innerbib\@empty
\bibitem [{\citenamefont {Hewson}(1993)}]{Hewson1993}%
  \BibitemOpen
  \bibfield  {author} {\bibinfo {author} {\bibfnamefont {A.~C.}\ \bibnamefont {Hewson}},\ }\href@noop {} {\emph {\bibinfo {title} {The {Kondo} problem to heavy fermions}}}\ (\bibinfo  {publisher} {Cambridge University Press},\ \bibinfo {year} {1993})\BibitemShut {NoStop}%
\bibitem [{\citenamefont {Krellner}\ \emph {et~al.}(2007)\citenamefont {Krellner}, \citenamefont {Kini}, \citenamefont {Br\"uning}, \citenamefont {Koch}, \citenamefont {Rosner}, \citenamefont {Nicklas}, \citenamefont {Baenitz},\ and\ \citenamefont {Geibel}}]{Krellner2007}%
  \BibitemOpen
  \bibfield  {author} {\bibinfo {author} {\bibfnamefont {C.}~\bibnamefont {Krellner}}, \bibinfo {author} {\bibfnamefont {N.~S.}\ \bibnamefont {Kini}}, \bibinfo {author} {\bibfnamefont {E.~M.}\ \bibnamefont {Br\"uning}}, \bibinfo {author} {\bibfnamefont {K.}~\bibnamefont {Koch}}, \bibinfo {author} {\bibfnamefont {H.}~\bibnamefont {Rosner}}, \bibinfo {author} {\bibfnamefont {M.}~\bibnamefont {Nicklas}}, \bibinfo {author} {\bibfnamefont {M.}~\bibnamefont {Baenitz}},\ and\ \bibinfo {author} {\bibfnamefont {C.}~\bibnamefont {Geibel}},\ }\bibfield  {title} {\bibinfo {title} {{CeRuPO}: A rare example of a ferromagnetic {Kondo} lattice},\ }\href {https://doi.org/10.1103/PhysRevB.76.104418} {\bibfield  {journal} {\bibinfo  {journal} {Phys. Rev. B}\ }\textbf {\bibinfo {volume} {76}},\ \bibinfo {pages} {104418} (\bibinfo {year} {2007})}\BibitemShut {NoStop}%
\bibitem [{\citenamefont {Baumbach}\ \emph {et~al.}(2012)\citenamefont {Baumbach}, \citenamefont {Chudo}, \citenamefont {Yasuoka}, \citenamefont {Ronning}, \citenamefont {Bauer},\ and\ \citenamefont {Thompson}}]{Baumbach2012}%
  \BibitemOpen
  \bibfield  {author} {\bibinfo {author} {\bibfnamefont {R.~E.}\ \bibnamefont {Baumbach}}, \bibinfo {author} {\bibfnamefont {H.}~\bibnamefont {Chudo}}, \bibinfo {author} {\bibfnamefont {H.}~\bibnamefont {Yasuoka}}, \bibinfo {author} {\bibfnamefont {F.}~\bibnamefont {Ronning}}, \bibinfo {author} {\bibfnamefont {E.~D.}\ \bibnamefont {Bauer}},\ and\ \bibinfo {author} {\bibfnamefont {J.~D.}\ \bibnamefont {Thompson}},\ }\bibfield  {title} {\bibinfo {title} {{CeRu${}_{2}$Al${}_{2}$B}: A local-moment $4f$ magnet with a complex {$T$-$H$} phase diagram},\ }\href {https://doi.org/10.1103/PhysRevB.85.094422} {\bibfield  {journal} {\bibinfo  {journal} {Phys. Rev. B}\ }\textbf {\bibinfo {volume} {85}},\ \bibinfo {pages} {094422} (\bibinfo {year} {2012})}\BibitemShut {NoStop}%
\bibitem [{\citenamefont {J.}\ \emph {et~al.}(2005)\citenamefont {J.}, \citenamefont {Fontes}, \citenamefont {Alvarenga}, \citenamefont {Baggio-Saitovitch}, \citenamefont {Burghardt}, \citenamefont {Eichler},\ and\ \citenamefont {Continentino}}]{Larrea2005}%
  \BibitemOpen
  \bibfield  {author} {\bibinfo {author} {\bibfnamefont {J.~L.}\ \bibnamefont {J.}}, \bibinfo {author} {\bibfnamefont {M.~B.}\ \bibnamefont {Fontes}}, \bibinfo {author} {\bibfnamefont {A.~D.}\ \bibnamefont {Alvarenga}}, \bibinfo {author} {\bibfnamefont {E.~M.}\ \bibnamefont {Baggio-Saitovitch}}, \bibinfo {author} {\bibfnamefont {T.}~\bibnamefont {Burghardt}}, \bibinfo {author} {\bibfnamefont {A.}~\bibnamefont {Eichler}},\ and\ \bibinfo {author} {\bibfnamefont {M.~A.}\ \bibnamefont {Continentino}},\ }\bibfield  {title} {\bibinfo {title} {Quantum critical behavior in a \ce{CePt} ferromagnetic {Kondo} lattice},\ }\href {https://doi.org/10.1103/PhysRevB.72.035129} {\bibfield  {journal} {\bibinfo  {journal} {Phys. Rev. B}\ }\textbf {\bibinfo {volume} {72}},\ \bibinfo {pages} {035129} (\bibinfo {year} {2005})}\BibitemShut {NoStop}%
\bibitem [{\citenamefont {Brando}\ \emph {et~al.}(2016)\citenamefont {Brando}, \citenamefont {Belitz}, \citenamefont {Grosche},\ and\ \citenamefont {Kirkpatrick}}]{Brando2016}%
  \BibitemOpen
  \bibfield  {author} {\bibinfo {author} {\bibfnamefont {M.}~\bibnamefont {Brando}}, \bibinfo {author} {\bibfnamefont {D.}~\bibnamefont {Belitz}}, \bibinfo {author} {\bibfnamefont {F.~M.}\ \bibnamefont {Grosche}},\ and\ \bibinfo {author} {\bibfnamefont {T.~R.}\ \bibnamefont {Kirkpatrick}},\ }\bibfield  {title} {\bibinfo {title} {Metallic quantum ferromagnets},\ }\href {https://doi.org/10.1103/RevModPhys.88.025006} {\bibfield  {journal} {\bibinfo  {journal} {Rev. Mod. Phys.}\ }\textbf {\bibinfo {volume} {88}},\ \bibinfo {pages} {025006} (\bibinfo {year} {2016})}\BibitemShut {NoStop}%
\bibitem [{\citenamefont {Huxley}\ \emph {et~al.}(2001)\citenamefont {Huxley}, \citenamefont {Sheikin}, \citenamefont {Ressouche}, \citenamefont {Kernavanois}, \citenamefont {Braithwaite}, \citenamefont {Calemczuk},\ and\ \citenamefont {Flouquet}}]{Huxley2001}%
  \BibitemOpen
  \bibfield  {author} {\bibinfo {author} {\bibfnamefont {A.}~\bibnamefont {Huxley}}, \bibinfo {author} {\bibfnamefont {I.}~\bibnamefont {Sheikin}}, \bibinfo {author} {\bibfnamefont {E.}~\bibnamefont {Ressouche}}, \bibinfo {author} {\bibfnamefont {N.}~\bibnamefont {Kernavanois}}, \bibinfo {author} {\bibfnamefont {D.}~\bibnamefont {Braithwaite}}, \bibinfo {author} {\bibfnamefont {R.}~\bibnamefont {Calemczuk}},\ and\ \bibinfo {author} {\bibfnamefont {J.}~\bibnamefont {Flouquet}},\ }\bibfield  {title} {\bibinfo {title} {{\ce{UGe2}} a ferromagnetic spin-triplet superconductor},\ }\href {https://doi.org/10.1103/PhysRevB.63.144519} {\bibfield  {journal} {\bibinfo  {journal} {Phys. Rev. B}\ }\textbf {\bibinfo {volume} {63}},\ \bibinfo {pages} {144519} (\bibinfo {year} {2001})}\BibitemShut {NoStop}%
\bibitem [{\citenamefont {Aoki}\ \emph {et~al.}(2011)\citenamefont {Aoki}, \citenamefont {Combier}, \citenamefont {Taufour}, \citenamefont {D.~Matsuda}, \citenamefont {Knebel}, \citenamefont {Kotegawa},\ and\ \citenamefont {Flouquet}}]{Aoki2011}%
  \BibitemOpen
  \bibfield  {author} {\bibinfo {author} {\bibfnamefont {D.}~\bibnamefont {Aoki}}, \bibinfo {author} {\bibfnamefont {T.}~\bibnamefont {Combier}}, \bibinfo {author} {\bibfnamefont {V.}~\bibnamefont {Taufour}}, \bibinfo {author} {\bibfnamefont {T.}~\bibnamefont {D.~Matsuda}}, \bibinfo {author} {\bibfnamefont {G.}~\bibnamefont {Knebel}}, \bibinfo {author} {\bibfnamefont {H.}~\bibnamefont {Kotegawa}},\ and\ \bibinfo {author} {\bibfnamefont {J.}~\bibnamefont {Flouquet}},\ }\bibfield  {title} {\bibinfo {title} {Ferromagnetic quantum critical endpoint in \ce{UCoAl}},\ }\href {https://doi.org/10.1143/JPSJ.80.094711} {\bibfield  {journal} {\bibinfo  {journal} {J. Phys. Soc. Jpn.}\ }\textbf {\bibinfo {volume} {80}},\ \bibinfo {pages} {094711} (\bibinfo {year} {2011})}\BibitemShut {NoStop}%
\bibitem [{\citenamefont {Matsuoka}\ \emph {et~al.}(2015)\citenamefont {Matsuoka}, \citenamefont {Hondo}, \citenamefont {Fujii}, \citenamefont {Oshima}, \citenamefont {Sugawara}, \citenamefont {Sakurai}, \citenamefont {Ohta}, \citenamefont {Kneidinger}, \citenamefont {Salamakha}, \citenamefont {Michor},\ and\ \citenamefont {Bauer}}]{Matsuoka2015}%
  \BibitemOpen
  \bibfield  {author} {\bibinfo {author} {\bibfnamefont {E.}~\bibnamefont {Matsuoka}}, \bibinfo {author} {\bibfnamefont {C.}~\bibnamefont {Hondo}}, \bibinfo {author} {\bibfnamefont {T.}~\bibnamefont {Fujii}}, \bibinfo {author} {\bibfnamefont {A.}~\bibnamefont {Oshima}}, \bibinfo {author} {\bibfnamefont {H.}~\bibnamefont {Sugawara}}, \bibinfo {author} {\bibfnamefont {T.}~\bibnamefont {Sakurai}}, \bibinfo {author} {\bibfnamefont {H.}~\bibnamefont {Ohta}}, \bibinfo {author} {\bibfnamefont {F.}~\bibnamefont {Kneidinger}}, \bibinfo {author} {\bibfnamefont {L.}~\bibnamefont {Salamakha}}, \bibinfo {author} {\bibfnamefont {H.}~\bibnamefont {Michor}},\ and\ \bibinfo {author} {\bibfnamefont {E.}~\bibnamefont {Bauer}},\ }\bibfield  {title} {\bibinfo {title} {Ferromagnetic transition at {2.5\,K} in the hexagonal {Kondo}-lattice compound \ce{CeRh6Ge4}},\ }\href {https://doi.org/10.7566/JPSJ.84.073704} {\bibfield  {journal} {\bibinfo  {journal} {J. Phys. Soc. Jpn.}\ }\textbf {\bibinfo {volume} {84}},\ \bibinfo {pages}
  {073704} (\bibinfo {year} {2015})}\BibitemShut {NoStop}%
\bibitem [{\citenamefont {Kotegawa}\ \emph {et~al.}(2019)\citenamefont {Kotegawa}, \citenamefont {Matsuoka}, \citenamefont {Uga}, \citenamefont {Takemura}, \citenamefont {Manago}, \citenamefont {Chikuchi}, \citenamefont {Sugawara}, \citenamefont {Tou},\ and\ \citenamefont {Harima}}]{Kotegawa2019}%
  \BibitemOpen
  \bibfield  {author} {\bibinfo {author} {\bibfnamefont {H.}~\bibnamefont {Kotegawa}}, \bibinfo {author} {\bibfnamefont {E.}~\bibnamefont {Matsuoka}}, \bibinfo {author} {\bibfnamefont {T.}~\bibnamefont {Uga}}, \bibinfo {author} {\bibfnamefont {M.}~\bibnamefont {Takemura}}, \bibinfo {author} {\bibfnamefont {M.}~\bibnamefont {Manago}}, \bibinfo {author} {\bibfnamefont {N.}~\bibnamefont {Chikuchi}}, \bibinfo {author} {\bibfnamefont {H.}~\bibnamefont {Sugawara}}, \bibinfo {author} {\bibfnamefont {H.}~\bibnamefont {Tou}},\ and\ \bibinfo {author} {\bibfnamefont {H.}~\bibnamefont {Harima}},\ }\bibfield  {title} {\bibinfo {title} {Indication of ferromagnetic quantum critical point in {Kondo} lattice \ce{CeRh6Ge4}},\ }\href {https://doi.org/10.7566/JPSJ.88.093702} {\bibfield  {journal} {\bibinfo  {journal} {J. Phys. Soc. Jpn.}\ }\textbf {\bibinfo {volume} {88}},\ \bibinfo {pages} {093702} (\bibinfo {year} {2019})}\BibitemShut {NoStop}%
\bibitem [{\citenamefont {Shen}\ \emph {et~al.}(2020)\citenamefont {Shen}, \citenamefont {Zhang}, \citenamefont {Komijani}, \citenamefont {Nicklas}, \citenamefont {Borth}, \citenamefont {Wang}, \citenamefont {Chen}, \citenamefont {Nie}, \citenamefont {Li}, \citenamefont {Lu}, \citenamefont {Lee}, \citenamefont {Smidman}, \citenamefont {Steglich}, \citenamefont {Coleman},\ and\ \citenamefont {Yuan}}]{Shen2020}%
  \BibitemOpen
  \bibfield  {author} {\bibinfo {author} {\bibfnamefont {B.}~\bibnamefont {Shen}}, \bibinfo {author} {\bibfnamefont {Y.}~\bibnamefont {Zhang}}, \bibinfo {author} {\bibfnamefont {Y.}~\bibnamefont {Komijani}}, \bibinfo {author} {\bibfnamefont {M.}~\bibnamefont {Nicklas}}, \bibinfo {author} {\bibfnamefont {R.}~\bibnamefont {Borth}}, \bibinfo {author} {\bibfnamefont {A.}~\bibnamefont {Wang}}, \bibinfo {author} {\bibfnamefont {Y.}~\bibnamefont {Chen}}, \bibinfo {author} {\bibfnamefont {Z.}~\bibnamefont {Nie}}, \bibinfo {author} {\bibfnamefont {R.}~\bibnamefont {Li}}, \bibinfo {author} {\bibfnamefont {X.}~\bibnamefont {Lu}}, \bibinfo {author} {\bibfnamefont {H.}~\bibnamefont {Lee}}, \bibinfo {author} {\bibfnamefont {M.}~\bibnamefont {Smidman}}, \bibinfo {author} {\bibfnamefont {F.}~\bibnamefont {Steglich}}, \bibinfo {author} {\bibfnamefont {P.}~\bibnamefont {Coleman}},\ and\ \bibinfo {author} {\bibfnamefont {H.}~\bibnamefont {Yuan}},\ }\bibfield  {title} {\bibinfo {title} {Strange-metal behaviour in a pure
  ferromagnetic {Kondo} lattice},\ }\href {https://doi.org/10.1038/s41586-020-2052-z} {\bibfield  {journal} {\bibinfo  {journal} {Nature}\ }\textbf {\bibinfo {volume} {579}},\ \bibinfo {pages} {51} (\bibinfo {year} {2020})}\BibitemShut {NoStop}%
\bibitem [{\citenamefont {Kirkpatrick}\ and\ \citenamefont {Belitz}(2020)}]{Kirkpatrick2020}%
  \BibitemOpen
  \bibfield  {author} {\bibinfo {author} {\bibfnamefont {T.~R.}\ \bibnamefont {Kirkpatrick}}\ and\ \bibinfo {author} {\bibfnamefont {D.}~\bibnamefont {Belitz}},\ }\bibfield  {title} {\bibinfo {title} {Ferromagnetic quantum critical point in noncentrosymmetric systems},\ }\href {https://doi.org/10.1103/PhysRevLett.124.147201} {\bibfield  {journal} {\bibinfo  {journal} {Phys. Rev. Lett.}\ }\textbf {\bibinfo {volume} {124}},\ \bibinfo {pages} {147201} (\bibinfo {year} {2020})}\BibitemShut {NoStop}%
\bibitem [{\citenamefont {Miserev}\ \emph {et~al.}(2022)\citenamefont {Miserev}, \citenamefont {Loss},\ and\ \citenamefont {Klinovaja}}]{Miserev2022}%
  \BibitemOpen
  \bibfield  {author} {\bibinfo {author} {\bibfnamefont {D.}~\bibnamefont {Miserev}}, \bibinfo {author} {\bibfnamefont {D.}~\bibnamefont {Loss}},\ and\ \bibinfo {author} {\bibfnamefont {J.}~\bibnamefont {Klinovaja}},\ }\bibfield  {title} {\bibinfo {title} {Instability of the ferromagnetic quantum critical point and symmetry of the ferromagnetic ground state in two-dimensional and three-dimensional electron gases with arbitrary spin-orbit splitting},\ }\href {https://doi.org/10.1103/PhysRevB.106.134417} {\bibfield  {journal} {\bibinfo  {journal} {Phys. Rev. B}\ }\textbf {\bibinfo {volume} {106}},\ \bibinfo {pages} {134417} (\bibinfo {year} {2022})}\BibitemShut {NoStop}%
\bibitem [{\citenamefont {Shin}\ \emph {et~al.}(2024)\citenamefont {Shin}, \citenamefont {Ramires}, \citenamefont {Pomjakushin}, \citenamefont {Plokhikh},\ and\ \citenamefont {Pomjakushina}}]{Shin2024}%
  \BibitemOpen
  \bibfield  {author} {\bibinfo {author} {\bibfnamefont {S.}~\bibnamefont {Shin}}, \bibinfo {author} {\bibfnamefont {A.}~\bibnamefont {Ramires}}, \bibinfo {author} {\bibfnamefont {V.}~\bibnamefont {Pomjakushin}}, \bibinfo {author} {\bibfnamefont {I.}~\bibnamefont {Plokhikh}},\ and\ \bibinfo {author} {\bibfnamefont {E.}~\bibnamefont {Pomjakushina}},\ }\bibfield  {title} {\bibinfo {title} {Ferromagnetic quantum critical point protected by nonsymmorphic symmetry in a {Kondo} metal},\ }\href {https://doi.org/10.1038/s41467-024-52720-9} {\bibfield  {journal} {\bibinfo  {journal} {Nat. Commun.}\ }\textbf {\bibinfo {volume} {15}},\ \bibinfo {pages} {8423} (\bibinfo {year} {2024})}\BibitemShut {NoStop}%
\bibitem [{\citenamefont {Shu}\ \emph {et~al.}(2021)\citenamefont {Shu}, \citenamefont {Adroja}, \citenamefont {Hillier}, \citenamefont {Zhang}, \citenamefont {Chen}, \citenamefont {Shen}, \citenamefont {Orlandi}, \citenamefont {Walker}, \citenamefont {Liu}, \citenamefont {Cao}, \citenamefont {Steglich}, \citenamefont {Yuan},\ and\ \citenamefont {Smidman}}]{Shu2021}%
  \BibitemOpen
  \bibfield  {author} {\bibinfo {author} {\bibfnamefont {J.~W.}\ \bibnamefont {Shu}}, \bibinfo {author} {\bibfnamefont {D.~T.}\ \bibnamefont {Adroja}}, \bibinfo {author} {\bibfnamefont {A.~D.}\ \bibnamefont {Hillier}}, \bibinfo {author} {\bibfnamefont {Y.~J.}\ \bibnamefont {Zhang}}, \bibinfo {author} {\bibfnamefont {Y.~X.}\ \bibnamefont {Chen}}, \bibinfo {author} {\bibfnamefont {B.}~\bibnamefont {Shen}}, \bibinfo {author} {\bibfnamefont {F.}~\bibnamefont {Orlandi}}, \bibinfo {author} {\bibfnamefont {H.~C.}\ \bibnamefont {Walker}}, \bibinfo {author} {\bibfnamefont {Y.}~\bibnamefont {Liu}}, \bibinfo {author} {\bibfnamefont {C.}~\bibnamefont {Cao}}, \bibinfo {author} {\bibfnamefont {F.}~\bibnamefont {Steglich}}, \bibinfo {author} {\bibfnamefont {H.~Q.}\ \bibnamefont {Yuan}},\ and\ \bibinfo {author} {\bibfnamefont {M.}~\bibnamefont {Smidman}},\ }\bibfield  {title} {\bibinfo {title} {Magnetic order and crystalline electric field excitations of the quantum critical heavy-fermion ferromagnet \ce{CeRh6Ge4}},\ }\href
  {https://doi.org/10.1103/PhysRevB.104.L140411} {\bibfield  {journal} {\bibinfo  {journal} {Phys. Rev. B}\ }\textbf {\bibinfo {volume} {104}},\ \bibinfo {pages} {L140411} (\bibinfo {year} {2021})}\BibitemShut {NoStop}%
\bibitem [{\citenamefont {Vo{\ss}inkel}\ \emph {et~al.}(2012)\citenamefont {Vo{\ss}inkel}, \citenamefont {Niehaus}, \citenamefont {Rodewald},\ and\ \citenamefont {P\"{o}ttgen}}]{Vossinkel2012}%
  \BibitemOpen
  \bibfield  {author} {\bibinfo {author} {\bibfnamefont {D.}~\bibnamefont {Vo{\ss}inkel}}, \bibinfo {author} {\bibfnamefont {O.}~\bibnamefont {Niehaus}}, \bibinfo {author} {\bibfnamefont {U.~C.}\ \bibnamefont {Rodewald}},\ and\ \bibinfo {author} {\bibfnamefont {R.}~\bibnamefont {P\"{o}ttgen}},\ }\bibfield  {title} {\bibinfo {title} {Bismuth flux growth of \ce{CeRh6Ge4} and \ce{CeRh2Ge2} single crystals},\ }\href {https://doi.org/doi:10.5560/znb.2012-0265} {\bibfield  {journal} {\bibinfo  {journal} {Z. Naturforsch. B}\ }\textbf {\bibinfo {volume} {67}},\ \bibinfo {pages} {1241} (\bibinfo {year} {2012})}\BibitemShut {NoStop}%
\bibitem [{\citenamefont {Wu}\ \emph {et~al.}(2021)\citenamefont {Wu}, \citenamefont {Zhang}, \citenamefont {Du}, \citenamefont {Shen}, \citenamefont {Zheng}, \citenamefont {Fang}, \citenamefont {Smidman}, \citenamefont {Cao}, \citenamefont {Steglich}, \citenamefont {Yuan}, \citenamefont {Denlinger},\ and\ \citenamefont {Liu}}]{Wu2021}%
  \BibitemOpen
  \bibfield  {author} {\bibinfo {author} {\bibfnamefont {Y.}~\bibnamefont {Wu}}, \bibinfo {author} {\bibfnamefont {Y.}~\bibnamefont {Zhang}}, \bibinfo {author} {\bibfnamefont {F.}~\bibnamefont {Du}}, \bibinfo {author} {\bibfnamefont {B.}~\bibnamefont {Shen}}, \bibinfo {author} {\bibfnamefont {H.}~\bibnamefont {Zheng}}, \bibinfo {author} {\bibfnamefont {Y.}~\bibnamefont {Fang}}, \bibinfo {author} {\bibfnamefont {M.}~\bibnamefont {Smidman}}, \bibinfo {author} {\bibfnamefont {C.}~\bibnamefont {Cao}}, \bibinfo {author} {\bibfnamefont {F.}~\bibnamefont {Steglich}}, \bibinfo {author} {\bibfnamefont {H.}~\bibnamefont {Yuan}}, \bibinfo {author} {\bibfnamefont {J.~D.}\ \bibnamefont {Denlinger}},\ and\ \bibinfo {author} {\bibfnamefont {Y.}~\bibnamefont {Liu}},\ }\bibfield  {title} {\bibinfo {title} {Anisotropic $c\ensuremath{-}f$ hybridization in the ferromagnetic quantum critical metal \ce{CeRh6Ge4}},\ }\href {https://doi.org/10.1103/PhysRevLett.126.216406} {\bibfield  {journal} {\bibinfo  {journal} {Phys. Rev.
  Lett.}\ }\textbf {\bibinfo {volume} {126}},\ \bibinfo {pages} {216406} (\bibinfo {year} {2021})}\BibitemShut {NoStop}%
\bibitem [{\citenamefont {Wang}\ \emph {et~al.}(2021)\citenamefont {Wang}, \citenamefont {Du}, \citenamefont {Zhang}, \citenamefont {Graf}, \citenamefont {Shen}, \citenamefont {Chen}, \citenamefont {Liu}, \citenamefont {Smidman}, \citenamefont {Cao}, \citenamefont {Steglich},\ and\ \citenamefont {Yuan}}]{Wang2021}%
  \BibitemOpen
  \bibfield  {author} {\bibinfo {author} {\bibfnamefont {A.}~\bibnamefont {Wang}}, \bibinfo {author} {\bibfnamefont {F.}~\bibnamefont {Du}}, \bibinfo {author} {\bibfnamefont {Y.}~\bibnamefont {Zhang}}, \bibinfo {author} {\bibfnamefont {D.}~\bibnamefont {Graf}}, \bibinfo {author} {\bibfnamefont {B.}~\bibnamefont {Shen}}, \bibinfo {author} {\bibfnamefont {Y.}~\bibnamefont {Chen}}, \bibinfo {author} {\bibfnamefont {Y.}~\bibnamefont {Liu}}, \bibinfo {author} {\bibfnamefont {M.}~\bibnamefont {Smidman}}, \bibinfo {author} {\bibfnamefont {C.}~\bibnamefont {Cao}}, \bibinfo {author} {\bibfnamefont {F.}~\bibnamefont {Steglich}},\ and\ \bibinfo {author} {\bibfnamefont {H.}~\bibnamefont {Yuan}},\ }\bibfield  {title} {\bibinfo {title} {Localized 4f-electrons in the quantum critical heavy fermion ferromagnet \ce{CeRh6Ge4}},\ }\href {https://doi.org/https://doi.org/10.1016/j.scib.2021.03.006} {\bibfield  {journal} {\bibinfo  {journal} {Science Bulletin}\ }\textbf {\bibinfo {volume} {66}},\ \bibinfo {pages} {1389} (\bibinfo
  {year} {2021})}\BibitemShut {NoStop}%
\bibitem [{\citenamefont {Xu}\ \emph {et~al.}(2021)\citenamefont {Xu}, \citenamefont {Su}, \citenamefont {Kumar}, \citenamefont {Luo}, \citenamefont {Nie}, \citenamefont {Wang}, \citenamefont {Du}, \citenamefont {Li}, \citenamefont {Smidman},\ and\ \citenamefont {Yuan}}]{Xu2021}%
  \BibitemOpen
  \bibfield  {author} {\bibinfo {author} {\bibfnamefont {J.-C.}\ \bibnamefont {Xu}}, \bibinfo {author} {\bibfnamefont {H.}~\bibnamefont {Su}}, \bibinfo {author} {\bibfnamefont {R.}~\bibnamefont {Kumar}}, \bibinfo {author} {\bibfnamefont {S.-S.}\ \bibnamefont {Luo}}, \bibinfo {author} {\bibfnamefont {Z.-Y.}\ \bibnamefont {Nie}}, \bibinfo {author} {\bibfnamefont {A.}~\bibnamefont {Wang}}, \bibinfo {author} {\bibfnamefont {F.}~\bibnamefont {Du}}, \bibinfo {author} {\bibfnamefont {R.}~\bibnamefont {Li}}, \bibinfo {author} {\bibfnamefont {M.}~\bibnamefont {Smidman}},\ and\ \bibinfo {author} {\bibfnamefont {H.-Q.}\ \bibnamefont {Yuan}},\ }\bibfield  {title} {\bibinfo {title} {Ce-site dilution in the ferromagnetic kondo lattice \ce{CeRh6Ge4}},\ }\href {https://doi.org/10.1088/0256-307X/38/8/087101} {\bibfield  {journal} {\bibinfo  {journal} {Chin. Phys. Lett.}\ }\textbf {\bibinfo {volume} {38}},\ \bibinfo {pages} {087101} (\bibinfo {year} {2021})}\BibitemShut {NoStop}%
\bibitem [{\citenamefont {Zhang}\ \emph {et~al.}(2022)\citenamefont {Zhang}, \citenamefont {Nie}, \citenamefont {Li}, \citenamefont {Li}, \citenamefont {Yang}, \citenamefont {Shen}, \citenamefont {Chen}, \citenamefont {Du}, \citenamefont {Luo}, \citenamefont {Su}, \citenamefont {Shi}, \citenamefont {Wang}, \citenamefont {Nicklas}, \citenamefont {Steglich}, \citenamefont {Smidman},\ and\ \citenamefont {Yuan}}]{Zhang2022}%
  \BibitemOpen
  \bibfield  {author} {\bibinfo {author} {\bibfnamefont {Y.~J.}\ \bibnamefont {Zhang}}, \bibinfo {author} {\bibfnamefont {Z.~Y.}\ \bibnamefont {Nie}}, \bibinfo {author} {\bibfnamefont {R.}~\bibnamefont {Li}}, \bibinfo {author} {\bibfnamefont {Y.~C.}\ \bibnamefont {Li}}, \bibinfo {author} {\bibfnamefont {D.~L.}\ \bibnamefont {Yang}}, \bibinfo {author} {\bibfnamefont {B.}~\bibnamefont {Shen}}, \bibinfo {author} {\bibfnamefont {Y.}~\bibnamefont {Chen}}, \bibinfo {author} {\bibfnamefont {F.}~\bibnamefont {Du}}, \bibinfo {author} {\bibfnamefont {S.~S.}\ \bibnamefont {Luo}}, \bibinfo {author} {\bibfnamefont {H.}~\bibnamefont {Su}}, \bibinfo {author} {\bibfnamefont {R.}~\bibnamefont {Shi}}, \bibinfo {author} {\bibfnamefont {S.~Y.}\ \bibnamefont {Wang}}, \bibinfo {author} {\bibfnamefont {M.}~\bibnamefont {Nicklas}}, \bibinfo {author} {\bibfnamefont {F.}~\bibnamefont {Steglich}}, \bibinfo {author} {\bibfnamefont {M.}~\bibnamefont {Smidman}},\ and\ \bibinfo {author} {\bibfnamefont {H.~Q.}\ \bibnamefont {Yuan}},\
  }\bibfield  {title} {\bibinfo {title} {Suppression of ferromagnetism and influence of disorder in silicon-substituted \ce{CeRh6Ge4}},\ }\href {https://doi.org/10.1103/PhysRevB.106.054409} {\bibfield  {journal} {\bibinfo  {journal} {Phys. Rev. B}\ }\textbf {\bibinfo {volume} {106}},\ \bibinfo {pages} {054409} (\bibinfo {year} {2022})}\BibitemShut {NoStop}%
\bibitem [{\citenamefont {Thomas}\ \emph {et~al.}(2024)\citenamefont {Thomas}, \citenamefont {Seo}, \citenamefont {Asaba}, \citenamefont {Ronning}, \citenamefont {Rosa}, \citenamefont {Bauer},\ and\ \citenamefont {Thompson}}]{Thomas2024}%
  \BibitemOpen
  \bibfield  {author} {\bibinfo {author} {\bibfnamefont {S.~M.}\ \bibnamefont {Thomas}}, \bibinfo {author} {\bibfnamefont {S.}~\bibnamefont {Seo}}, \bibinfo {author} {\bibfnamefont {T.}~\bibnamefont {Asaba}}, \bibinfo {author} {\bibfnamefont {F.}~\bibnamefont {Ronning}}, \bibinfo {author} {\bibfnamefont {P.~F.~S.}\ \bibnamefont {Rosa}}, \bibinfo {author} {\bibfnamefont {E.~D.}\ \bibnamefont {Bauer}},\ and\ \bibinfo {author} {\bibfnamefont {J.~D.}\ \bibnamefont {Thompson}},\ }\bibfield  {title} {\bibinfo {title} {Probing quantum criticality in ferromagnetic \ce{CeRh6Ge4}},\ }\href {https://doi.org/10.1103/PhysRevB.109.L121105} {\bibfield  {journal} {\bibinfo  {journal} {Phys. Rev. B}\ }\textbf {\bibinfo {volume} {109}},\ \bibinfo {pages} {L121105} (\bibinfo {year} {2024})}\BibitemShut {NoStop}%
\bibitem [{\citenamefont {Otsuki}\ \emph {et~al.}(2024)\citenamefont {Otsuki}, \citenamefont {Yoshimi}, \citenamefont {Shinaoka},\ and\ \citenamefont {Jeschke}}]{Otsuki2024}%
  \BibitemOpen
  \bibfield  {author} {\bibinfo {author} {\bibfnamefont {J.}~\bibnamefont {Otsuki}}, \bibinfo {author} {\bibfnamefont {K.}~\bibnamefont {Yoshimi}}, \bibinfo {author} {\bibfnamefont {H.}~\bibnamefont {Shinaoka}},\ and\ \bibinfo {author} {\bibfnamefont {H.~O.}\ \bibnamefont {Jeschke}},\ }\bibfield  {title} {\bibinfo {title} {Multipolar ordering from dynamical mean field theory with application to \ce{CeB6}},\ }\href {https://doi.org/10.1103/PhysRevB.110.035104} {\bibfield  {journal} {\bibinfo  {journal} {Phys. Rev. B}\ }\textbf {\bibinfo {volume} {110}},\ \bibinfo {pages} {035104} (\bibinfo {year} {2024})}\BibitemShut {NoStop}%
\bibitem [{\citenamefont {Koepernik}\ and\ \citenamefont {Eschrig}(1999)}]{Koepernik1999}%
  \BibitemOpen
  \bibfield  {author} {\bibinfo {author} {\bibfnamefont {K.}~\bibnamefont {Koepernik}}\ and\ \bibinfo {author} {\bibfnamefont {H.}~\bibnamefont {Eschrig}},\ }\bibfield  {title} {\bibinfo {title} {Full-potential nonorthogonal local-orbital minimum-basis band-structure scheme},\ }\href {https://doi.org/10.1103/PhysRevB.59.1743} {\bibfield  {journal} {\bibinfo  {journal} {Phys. Rev. B}\ }\textbf {\bibinfo {volume} {59}},\ \bibinfo {pages} {1743} (\bibinfo {year} {1999})}\BibitemShut {NoStop}%
\bibitem [{\citenamefont {Perdew}\ \emph {et~al.}(1996)\citenamefont {Perdew}, \citenamefont {Burke},\ and\ \citenamefont {Ernzerhof}}]{Perdew1996}%
  \BibitemOpen
  \bibfield  {author} {\bibinfo {author} {\bibfnamefont {J.~P.}\ \bibnamefont {Perdew}}, \bibinfo {author} {\bibfnamefont {K.}~\bibnamefont {Burke}},\ and\ \bibinfo {author} {\bibfnamefont {M.}~\bibnamefont {Ernzerhof}},\ }\bibfield  {title} {\bibinfo {title} {Generalized gradient approximation made simple},\ }\href {https://doi.org/10.1103/PhysRevLett.77.3865} {\bibfield  {journal} {\bibinfo  {journal} {Phys. Rev. Lett.}\ }\textbf {\bibinfo {volume} {77}},\ \bibinfo {pages} {3865} (\bibinfo {year} {1996})}\BibitemShut {NoStop}%
\bibitem [{\citenamefont {Eschrig}\ and\ \citenamefont {Koepernik}(2009)}]{Eschrig2009}%
  \BibitemOpen
  \bibfield  {author} {\bibinfo {author} {\bibfnamefont {H.}~\bibnamefont {Eschrig}}\ and\ \bibinfo {author} {\bibfnamefont {K.}~\bibnamefont {Koepernik}},\ }\bibfield  {title} {\bibinfo {title} {Tight-binding models for the iron-based superconductors},\ }\href {https://doi.org/10.1103/PhysRevB.80.104503} {\bibfield  {journal} {\bibinfo  {journal} {Phys. Rev. B}\ }\textbf {\bibinfo {volume} {80}},\ \bibinfo {pages} {104503} (\bibinfo {year} {2009})}\BibitemShut {NoStop}%
\bibitem [{\citenamefont {Koepernik}\ \emph {et~al.}(2023)\citenamefont {Koepernik}, \citenamefont {Janson}, \citenamefont {Sun},\ and\ \citenamefont {van~den Brink}}]{Koepernik2023}%
  \BibitemOpen
  \bibfield  {author} {\bibinfo {author} {\bibfnamefont {K.}~\bibnamefont {Koepernik}}, \bibinfo {author} {\bibfnamefont {O.}~\bibnamefont {Janson}}, \bibinfo {author} {\bibfnamefont {Y.}~\bibnamefont {Sun}},\ and\ \bibinfo {author} {\bibfnamefont {J.}~\bibnamefont {van~den Brink}},\ }\bibfield  {title} {\bibinfo {title} {Symmetry-conserving maximally projected {W}annier functions},\ }\href {https://doi.org/10.1103/PhysRevB.107.235135} {\bibfield  {journal} {\bibinfo  {journal} {Phys. Rev. B}\ }\textbf {\bibinfo {volume} {107}},\ \bibinfo {pages} {235135} (\bibinfo {year} {2023})}\BibitemShut {NoStop}%
\bibitem [{\citenamefont {Georges}\ \emph {et~al.}(1996)\citenamefont {Georges}, \citenamefont {Kotliar}, \citenamefont {Krauth},\ and\ \citenamefont {Rozenberg}}]{Georges1996}%
  \BibitemOpen
  \bibfield  {author} {\bibinfo {author} {\bibfnamefont {A.}~\bibnamefont {Georges}}, \bibinfo {author} {\bibfnamefont {G.}~\bibnamefont {Kotliar}}, \bibinfo {author} {\bibfnamefont {W.}~\bibnamefont {Krauth}},\ and\ \bibinfo {author} {\bibfnamefont {M.~J.}\ \bibnamefont {Rozenberg}},\ }\bibfield  {title} {\bibinfo {title} {Dynamical mean-field theory of strongly correlated fermion systems and the limit of infinite dimensions},\ }\href {https://doi.org/10.1103/RevModPhys.68.13} {\bibfield  {journal} {\bibinfo  {journal} {Rev. Mod. Phys.}\ }\textbf {\bibinfo {volume} {68}},\ \bibinfo {pages} {13} (\bibinfo {year} {1996})}\BibitemShut {NoStop}%
\bibitem [{\citenamefont {Held}\ \emph {et~al.}(2001)\citenamefont {Held}, \citenamefont {McMahan},\ and\ \citenamefont {Scalettar}}]{Held2001}%
  \BibitemOpen
  \bibfield  {author} {\bibinfo {author} {\bibfnamefont {K.}~\bibnamefont {Held}}, \bibinfo {author} {\bibfnamefont {A.~K.}\ \bibnamefont {McMahan}},\ and\ \bibinfo {author} {\bibfnamefont {R.~T.}\ \bibnamefont {Scalettar}},\ }\bibfield  {title} {\bibinfo {title} {{Cerium Volume Collapse: Results from the Merger of Dynamical Mean-Field Theory and Local Density Approximation}},\ }\href {https://doi.org/10.1103/PhysRevLett.87.276404} {\bibfield  {journal} {\bibinfo  {journal} {Phys. Rev. Lett.}\ }\textbf {\bibinfo {volume} {87}},\ \bibinfo {pages} {276404} (\bibinfo {year} {2001})}\BibitemShut {NoStop}%
\bibitem [{\citenamefont {Kotliar}\ \emph {et~al.}(2006)\citenamefont {Kotliar}, \citenamefont {Savrasov}, \citenamefont {Haule}, \citenamefont {Oudovenko}, \citenamefont {Parcollet},\ and\ \citenamefont {Marianetti}}]{Kotliar2006}%
  \BibitemOpen
  \bibfield  {author} {\bibinfo {author} {\bibfnamefont {G.}~\bibnamefont {Kotliar}}, \bibinfo {author} {\bibfnamefont {S.~Y.}\ \bibnamefont {Savrasov}}, \bibinfo {author} {\bibfnamefont {K.}~\bibnamefont {Haule}}, \bibinfo {author} {\bibfnamefont {V.~S.}\ \bibnamefont {Oudovenko}}, \bibinfo {author} {\bibfnamefont {O.}~\bibnamefont {Parcollet}},\ and\ \bibinfo {author} {\bibfnamefont {C.~A.}\ \bibnamefont {Marianetti}},\ }\bibfield  {title} {\bibinfo {title} {Electronic structure calculations with dynamical mean-field theory},\ }\href {https://doi.org/10.1103/RevModPhys.78.865} {\bibfield  {journal} {\bibinfo  {journal} {Rev. Mod. Phys.}\ }\textbf {\bibinfo {volume} {78}},\ \bibinfo {pages} {865} (\bibinfo {year} {2006})}\BibitemShut {NoStop}%
\bibitem [{\citenamefont {Locht}\ \emph {et~al.}(2016)\citenamefont {Locht}, \citenamefont {Kvashnin}, \citenamefont {Rodrigues}, \citenamefont {Pereiro}, \citenamefont {Bergman}, \citenamefont {Bergqvist}, \citenamefont {Lichtenstein}, \citenamefont {Katsnelson}, \citenamefont {Delin}, \citenamefont {Klautau}, \citenamefont {Johansson}, \citenamefont {Di~Marco},\ and\ \citenamefont {Eriksson}}]{Locht2016}%
  \BibitemOpen
  \bibfield  {author} {\bibinfo {author} {\bibfnamefont {I.~L.~M.}\ \bibnamefont {Locht}}, \bibinfo {author} {\bibfnamefont {Y.~O.}\ \bibnamefont {Kvashnin}}, \bibinfo {author} {\bibfnamefont {D.~C.~M.}\ \bibnamefont {Rodrigues}}, \bibinfo {author} {\bibfnamefont {M.}~\bibnamefont {Pereiro}}, \bibinfo {author} {\bibfnamefont {A.}~\bibnamefont {Bergman}}, \bibinfo {author} {\bibfnamefont {L.}~\bibnamefont {Bergqvist}}, \bibinfo {author} {\bibfnamefont {A.~I.}\ \bibnamefont {Lichtenstein}}, \bibinfo {author} {\bibfnamefont {M.~I.}\ \bibnamefont {Katsnelson}}, \bibinfo {author} {\bibfnamefont {A.}~\bibnamefont {Delin}}, \bibinfo {author} {\bibfnamefont {A.~B.}\ \bibnamefont {Klautau}}, \bibinfo {author} {\bibfnamefont {B.}~\bibnamefont {Johansson}}, \bibinfo {author} {\bibfnamefont {I.}~\bibnamefont {Di~Marco}},\ and\ \bibinfo {author} {\bibfnamefont {O.}~\bibnamefont {Eriksson}},\ }\bibfield  {title} {\bibinfo {title} {{Standard model of the rare earths analyzed from the Hubbard I approximation}},\ }\href
  {https://doi.org/10.1103/PhysRevB.94.085137} {\bibfield  {journal} {\bibinfo  {journal} {Phys. Rev. B}\ }\textbf {\bibinfo {volume} {94}},\ \bibinfo {pages} {085137} (\bibinfo {year} {2016})}\BibitemShut {NoStop}%
\bibitem [{\citenamefont {Herbst}\ \emph {et~al.}(1978)\citenamefont {Herbst}, \citenamefont {Watson},\ and\ \citenamefont {Wilkins}}]{Herbst1978}%
  \BibitemOpen
  \bibfield  {author} {\bibinfo {author} {\bibfnamefont {J.~F.}\ \bibnamefont {Herbst}}, \bibinfo {author} {\bibfnamefont {R.~E.}\ \bibnamefont {Watson}},\ and\ \bibinfo {author} {\bibfnamefont {J.~W.}\ \bibnamefont {Wilkins}},\ }\bibfield  {title} {\bibinfo {title} {Relativistic calculations of $4f$ excitation energies in the rare-earth metals: Further results},\ }\href {https://doi.org/10.1103/PhysRevB.17.3089} {\bibfield  {journal} {\bibinfo  {journal} {Phys. Rev. B}\ }\textbf {\bibinfo {volume} {17}},\ \bibinfo {pages} {3089} (\bibinfo {year} {1978})}\BibitemShut {NoStop}%
\bibitem [{\citenamefont {Lang}\ \emph {et~al.}(1981)\citenamefont {Lang}, \citenamefont {Baer},\ and\ \citenamefont {Cox}}]{Lang1981}%
  \BibitemOpen
  \bibfield  {author} {\bibinfo {author} {\bibfnamefont {J.~K.}\ \bibnamefont {Lang}}, \bibinfo {author} {\bibfnamefont {Y.}~\bibnamefont {Baer}},\ and\ \bibinfo {author} {\bibfnamefont {P.~A.}\ \bibnamefont {Cox}},\ }\bibfield  {title} {\bibinfo {title} {{Study of the 4f and valence band density of states in rare-earth metals. II. Experiment and results}},\ }\href {https://doi.org/10.1088/0305-4608/11/1/015} {\bibfield  {journal} {\bibinfo  {journal} {J. Phys. F: Met. Phys.}\ }\textbf {\bibinfo {volume} {11}},\ \bibinfo {pages} {121} (\bibinfo {year} {1981})}\BibitemShut {NoStop}%
\bibitem [{\citenamefont {Kusunose}\ \emph {et~al.}(2023)\citenamefont {Kusunose}, \citenamefont {Oiwa},\ and\ \citenamefont {Hayami}}]{Kusunose2023}%
  \BibitemOpen
  \bibfield  {author} {\bibinfo {author} {\bibfnamefont {H.}~\bibnamefont {Kusunose}}, \bibinfo {author} {\bibfnamefont {R.}~\bibnamefont {Oiwa}},\ and\ \bibinfo {author} {\bibfnamefont {S.}~\bibnamefont {Hayami}},\ }\bibfield  {title} {\bibinfo {title} {Symmetry-adapted modeling for molecules and crystals},\ }\href {https://doi.org/10.1103/PhysRevB.107.195118} {\bibfield  {journal} {\bibinfo  {journal} {Phys. Rev. B}\ }\textbf {\bibinfo {volume} {107}},\ \bibinfo {pages} {195118} (\bibinfo {year} {2023})}\BibitemShut {NoStop}%
\bibitem [{\citenamefont {Doniach}(1977)}]{Doniach1977}%
  \BibitemOpen
  \bibfield  {author} {\bibinfo {author} {\bibfnamefont {S.}~\bibnamefont {Doniach}},\ }\bibfield  {title} {\bibinfo {title} {The {Kondo} lattice and weak antiferromagnetism},\ }\href {https://doi.org/https://doi.org/10.1016/0378-4363(77)90190-5} {\bibfield  {journal} {\bibinfo  {journal} {Physica B+C}\ }\textbf {\bibinfo {volume} {91}},\ \bibinfo {pages} {231} (\bibinfo {year} {1977})}\BibitemShut {NoStop}%
\bibitem [{\citenamefont {Gull}\ \emph {et~al.}(2011)\citenamefont {Gull}, \citenamefont {Millis}, \citenamefont {Lichtenstein}, \citenamefont {Rubtsov}, \citenamefont {Troyer},\ and\ \citenamefont {Werner}}]{Gull2011}%
  \BibitemOpen
  \bibfield  {author} {\bibinfo {author} {\bibfnamefont {E.}~\bibnamefont {Gull}}, \bibinfo {author} {\bibfnamefont {A.~J.}\ \bibnamefont {Millis}}, \bibinfo {author} {\bibfnamefont {A.~I.}\ \bibnamefont {Lichtenstein}}, \bibinfo {author} {\bibfnamefont {A.~N.}\ \bibnamefont {Rubtsov}}, \bibinfo {author} {\bibfnamefont {M.}~\bibnamefont {Troyer}},\ and\ \bibinfo {author} {\bibfnamefont {P.}~\bibnamefont {Werner}},\ }\bibfield  {title} {\bibinfo {title} {Continuous-time {Monte Carlo} methods for quantum impurity models},\ }\href {https://doi.org/10.1103/RevModPhys.83.349} {\bibfield  {journal} {\bibinfo  {journal} {Rev. Mod. Phys.}\ }\textbf {\bibinfo {volume} {83}},\ \bibinfo {pages} {349} (\bibinfo {year} {2011})}\BibitemShut {NoStop}%
\bibitem [{\citenamefont {Anisimov}\ \emph {et~al.}(1997)\citenamefont {Anisimov}, \citenamefont {Aryasetiawan},\ and\ \citenamefont {Lichtenstein}}]{Anisimov1997}%
  \BibitemOpen
  \bibfield  {author} {\bibinfo {author} {\bibfnamefont {V.~I.}\ \bibnamefont {Anisimov}}, \bibinfo {author} {\bibfnamefont {F.}~\bibnamefont {Aryasetiawan}},\ and\ \bibinfo {author} {\bibfnamefont {A.~I.}\ \bibnamefont {Lichtenstein}},\ }\bibfield  {title} {\bibinfo {title} {{First-principles calculations of the electronic structure and spectra of strongly correlated systems: the LDA + U method}},\ }\href {https://doi.org/10.1088/0953-8984/9/4/002} {\bibfield  {journal} {\bibinfo  {journal} {J. Phys.: Condens. Matter}\ }\textbf {\bibinfo {volume} {9}},\ \bibinfo {pages} {767} (\bibinfo {year} {1997})}\BibitemShut {NoStop}%
\bibitem [{\citenamefont {Shinaoka}\ \emph {et~al.}(2020)\citenamefont {Shinaoka}, \citenamefont {Otsuki}, \citenamefont {Kawamura}, \citenamefont {Takemori},\ and\ \citenamefont {Yoshimi}}]{Shinaoka2021}%
  \BibitemOpen
  \bibfield  {author} {\bibinfo {author} {\bibfnamefont {H.}~\bibnamefont {Shinaoka}}, \bibinfo {author} {\bibfnamefont {J.}~\bibnamefont {Otsuki}}, \bibinfo {author} {\bibfnamefont {M.}~\bibnamefont {Kawamura}}, \bibinfo {author} {\bibfnamefont {N.}~\bibnamefont {Takemori}},\ and\ \bibinfo {author} {\bibfnamefont {K.}~\bibnamefont {Yoshimi}},\ }\bibfield  {title} {\bibinfo {title} {{DCore: Integrated DMFT software for correlated electrons}},\ }\href {https://doi.org/10.21468/SciPostPhys.10.5.117} {\bibfield  {journal} {\bibinfo  {journal} {SciPost Phys.}\ }\textbf {\bibinfo {volume} {10}},\ \bibinfo {pages} {117} (\bibinfo {year} {2020})}\BibitemShut {NoStop}%
\bibitem [{\citenamefont {Parcollet}\ \emph {et~al.}(2015)\citenamefont {Parcollet}, \citenamefont {Ferrero}, \citenamefont {Ayral}, \citenamefont {Hafermann}, \citenamefont {Krivenko}, \citenamefont {Messio},\ and\ \citenamefont {Seth}}]{Parcollet2015}%
  \BibitemOpen
  \bibfield  {author} {\bibinfo {author} {\bibfnamefont {O.}~\bibnamefont {Parcollet}}, \bibinfo {author} {\bibfnamefont {M.}~\bibnamefont {Ferrero}}, \bibinfo {author} {\bibfnamefont {T.}~\bibnamefont {Ayral}}, \bibinfo {author} {\bibfnamefont {H.}~\bibnamefont {Hafermann}}, \bibinfo {author} {\bibfnamefont {I.}~\bibnamefont {Krivenko}}, \bibinfo {author} {\bibfnamefont {L.}~\bibnamefont {Messio}},\ and\ \bibinfo {author} {\bibfnamefont {P.}~\bibnamefont {Seth}},\ }\bibfield  {title} {\bibinfo {title} {{TRIQS: A toolbox for research on interacting quantum systems}},\ }\href {https://doi.org/10.1016/j.cpc.2015.04.023} {\bibfield  {journal} {\bibinfo  {journal} {Comput. Phys. Commun.}\ }\textbf {\bibinfo {volume} {196}},\ \bibinfo {pages} {398} (\bibinfo {year} {2015})}\BibitemShut {NoStop}%
\bibitem [{\citenamefont {Aichhorn}\ \emph {et~al.}(2016)\citenamefont {Aichhorn}, \citenamefont {Pourovskii}, \citenamefont {Seth}, \citenamefont {Vildosola}, \citenamefont {Zingl}, \citenamefont {Peil}, \citenamefont {Deng}, \citenamefont {Mravlje}, \citenamefont {Kraberger}, \citenamefont {Martins}, \citenamefont {Ferrero},\ and\ \citenamefont {Parcollet}}]{Aichhorn2016}%
  \BibitemOpen
  \bibfield  {author} {\bibinfo {author} {\bibfnamefont {M.}~\bibnamefont {Aichhorn}}, \bibinfo {author} {\bibfnamefont {L.}~\bibnamefont {Pourovskii}}, \bibinfo {author} {\bibfnamefont {P.}~\bibnamefont {Seth}}, \bibinfo {author} {\bibfnamefont {V.}~\bibnamefont {Vildosola}}, \bibinfo {author} {\bibfnamefont {M.}~\bibnamefont {Zingl}}, \bibinfo {author} {\bibfnamefont {O.~E.}\ \bibnamefont {Peil}}, \bibinfo {author} {\bibfnamefont {X.}~\bibnamefont {Deng}}, \bibinfo {author} {\bibfnamefont {J.}~\bibnamefont {Mravlje}}, \bibinfo {author} {\bibfnamefont {G.~J.}\ \bibnamefont {Kraberger}}, \bibinfo {author} {\bibfnamefont {C.}~\bibnamefont {Martins}}, \bibinfo {author} {\bibfnamefont {M.}~\bibnamefont {Ferrero}},\ and\ \bibinfo {author} {\bibfnamefont {O.}~\bibnamefont {Parcollet}},\ }\bibfield  {title} {\bibinfo {title} {{TRIQS/DFTTools: A TRIQS application for ab initio calculations of correlated materials}},\ }\href {https://doi.org/10.1016/j.cpc.2016.03.014} {\bibfield  {journal} {\bibinfo  {journal}
  {Comput. Phys. Commun.}\ }\textbf {\bibinfo {volume} {204}},\ \bibinfo {pages} {200} (\bibinfo {year} {2016})}\BibitemShut {NoStop}%
\bibitem [{\citenamefont {Antipov}\ \emph {et~al.}(2017)\citenamefont {Antipov}, \citenamefont {Krivenko},\ and\ \citenamefont {Iskakov}}]{pomerol}%
  \BibitemOpen
  \bibfield  {author} {\bibinfo {author} {\bibfnamefont {A.~E.}\ \bibnamefont {Antipov}}, \bibinfo {author} {\bibfnamefont {I.}~\bibnamefont {Krivenko}},\ and\ \bibinfo {author} {\bibfnamefont {S.}~\bibnamefont {Iskakov}},\ }\href {https://doi.org/10.5281/zenodo.825870} {\bibinfo {title} {aeantipov/pomerol: 1.2}} (\bibinfo {year} {2017})\BibitemShut {NoStop}%
\bibitem [{\citenamefont {Otsuki}\ \emph {et~al.}(2019)\citenamefont {Otsuki}, \citenamefont {Yoshimi}, \citenamefont {Shinaoka},\ and\ \citenamefont {Nomura}}]{Otsuki2019}%
  \BibitemOpen
  \bibfield  {author} {\bibinfo {author} {\bibfnamefont {J.}~\bibnamefont {Otsuki}}, \bibinfo {author} {\bibfnamefont {K.}~\bibnamefont {Yoshimi}}, \bibinfo {author} {\bibfnamefont {H.}~\bibnamefont {Shinaoka}},\ and\ \bibinfo {author} {\bibfnamefont {Y.}~\bibnamefont {Nomura}},\ }\bibfield  {title} {\bibinfo {title} {Strong-coupling formula for momentum-dependent susceptibilities in dynamical mean-field theory},\ }\href {https://doi.org/10.1103/PhysRevB.99.165134} {\bibfield  {journal} {\bibinfo  {journal} {Phys. Rev. B}\ }\textbf {\bibinfo {volume} {99}},\ \bibinfo {pages} {165134} (\bibinfo {year} {2019})}\BibitemShut {NoStop}%
\end{thebibliography}%

\end{document}